\documentclass[10pt,openany,a4paper]{article}
\usepackage{geometry,color,framed}
\usepackage{amsmath,amssymb,mathtools,slashed,xcolor,mdframed}
\usepackage[vcentermath,enableskew]{youngtab}
\usepackage{cancel, pifont,multirow}
\usepackage{tikz}
\usetikzlibrary{positioning}
\usetikzlibrary{decorations.text}
\usepackage{amstext}
\usepackage{graphicx}
\usepackage[section]{placeins}
\usepackage{verbatim}
\usepackage{bm}
\usepackage{caption, subcaption, float}
\usepackage[colorlinks=true, 
            linkcolor=blue, citecolor=blue,
            menucolor = red,
            urlcolor=blue,linktoc=all]{hyperref}
\makeatletter

\newcommand{\beq}{\begin{eqnarray}}
\newcommand{\eeq}{\end{eqnarray}}
\newcommand{\non}{\nonumber\\}

\newcommand{\mb}{\mathbb}
\newcommand{\mf}{\mathbf}

\newcommand{\ben}{\begin{enumerate}}
\newcommand{\bei}{\begin{itemize}}
\newcommand{\eni}{\end{itemize}}
\newcommand{\enn}{\end{enumerate}}
\newcommand{\ra}{\rightarrow}
\newcommand{\rr}{\right}
\newcommand{\lf}{\left}

\newcommand{\xdownarrow}[1]{%
  {\left\downarrow\vbox to #1{}\right.\kern-\nulldelimiterspace}
}

\def\l{{\lambda}}

\def\s{{\sigma}}

\usepackage{tikz}
\usetikzlibrary{matrix,shapes,arrows,positioning,chains}
\tikzset{
	 connector/.style={
        -latex,
        font=\scriptsize
    },
    rectangle connector/.style={
        connector,
        to path={(\tikztostart) -- ++(#1,0pt) \tikztonodes |- (\tikztotarget) },
        pos=0.5
    },
    rectangle connector/.default=-2cm,
    straight connector/.style={
        connector,
        to path=--(\tikztotarget) \tikztonodes
    }
}

\makeatother
\title{Two-dimensional Hyperbolic RNN Neural Quantum State}
\author{H. L. Dao\footnote{espoirdujour1162@gmail.com}}
\date{\today}

\begin{document} 
\maketitle
\begin{abstract}
In the first part of this work, we construct the first type of two-dimensional (2D) hyperbolic neural quantum state (NQS) in the form of the Lorentz 2DRNN (Recurrent Neural Network) and benchmark its performance against the Euclidean 2DRNN in the paradigmatic $N\times N$ 2D Transverse Field Ising Model (2DTFIM) setting with different lattice sizes up to $N=12$ at different transverse magnetic field strengths. We find that hyperbolic Lorentz 2DRNN NQS robustly outperforms Euclidean 2DRNN NQS in our experiments when the system is at the phase transition point, where the physics can be described by a conformal field theory (CFT) - known via the AdS/CFT holographic duality to possess an underlying hyperbolic geometry. In the second part of this work, we benchmark the performances of recently introduced one-dimensional (1D) hyperbolic NQS ansatzes (including Poincar\'e RNN/GRU and Lorentz RNN/GRU) against their Euclidean versions in $N\times N$ 2DTFIM, converted to a one-dimensional sequence. The findings extend our previous results showing that 1D hyperbolic NQS robustly outperforms 1D Euclidean NQS, thanks to the combined effects of the induced $1^{\text{st}}$ and $N^{\text{th}}$ neighbor interaction hierarchy and the CFT physics at criticality. While studies on larger system sizes remain future work, this study serves as a proof-of-concept for the utility, effectiveness, and superior performance of 1D and 2D hyperbolic NQS ansatzes compared to conventional Euclidean NQS models in many-body quantum systems - particularly those exhibiting structural hierarchy, criticality, or a combination of both. 
\end{abstract}

\tableofcontents
\newpage
\section{Introduction}
Since the seminal works of \cite{1606-rbm}-\cite{1709-ann} that established the viability and utility of the simplest types of neural networks such as Restricted Boltzmann Machine and feedforward networks as wavefunction approximations in quantum many-body systems, steady progress has followed with ever-increasing complex architectures involving CNNs (Convolutional Neural Networks), RNNs (Recurrent Neural Networks) and transformers \cite{cnn-1807} - \cite{2306-transformer}.
In our recent works \cite{hld-hypnqs-25}, \cite{hld-hypnqs-26}, we introduced the first non-Euclidean neural quantum states in the form of the hyperbolic recurrent networks, including Poincar\'e RNN/GRU (Gated Recurrent Unit) and Lorentz RNN/GRU, whose underlying geometries are the Poincar\'e and Lorentz models of hyperbolic space instead of Euclidean space like all other NQS constructions that came before.  In these papers, we showed that these hyperbolic NQS could achieve superior performances compared to their Euclidean counterparts in various many-body quantum system settings that exhibit some hierarchical structure in the form of the different degrees of nearest neighbor interactions such as the Transverse Field Ising models (2DTFIM) as well as the Heisenberg $J_1J_2$ and Heisenberg $J_1J_2J_3$ models. In constructing and proposing these hyperbolic NQS, we were inspired by results from NLP (Natural Language Processing) \cite{ganea-1805}, \cite{hyprnn-lorentz-2105}, \cite{2010-hyprnn}, \cite{1901-vae} as well as graph-related tasks (graph embedding, link prediction, node classification, etc.) \cite{graph-classification}. These NLP works demonstrated the definitive outperformances delivered by different types of hyperbolic neural networks compared to their Euclidean versions in settings where hierarchical, tree-like structures exist in the training datasets. It is in these settings that the impressive performances by hyperbolic neural networks could be attributed to the fact that the exponential growth of hyperbolic space (area and volume)  allows for a very low distortion embedding of tree-like, hierarchical structures that is not possible with Euclidean space (whose area and volume only grow polynomially with distance) \cite{combinatorica-95}, \cite{krioukov-09}, \cite{krioukov-10},  \cite{sarkar-11}, \cite{sala-2018}.
\\\\
 While the work \cite{hld-hypnqs-26} enlarged the class of non-Euclidean hyperbolic NQS to include new constructions like Lorentz RNN/GRU and Poincar\'e RNN beyond the first hyperbolic NQS (the Poincar\'e GRU) that was introduced in \cite{hld-hypnqs-25}, the scope of the work \cite{hld-hypnqs-26} was restricted to the Heisenberg quantum systems. In \cite{hld-hypnqs-25}, we also studied the performances of Poincar\'e GRU in the $N\times N$ 2DTFIM context (up to $N=9$) and reported its notable outperformance over the Euclidean GRU NQS when the two dimensional problem is mapped to a one dimensional setup, the process of which introduces a hierarchical structure in the form of the first and $N^{th}$ neighbor interactions. In this work, we revisit the  problem of applying hyperbolic NQS to the 2D TFIM setting at different magnetic field strengths using larger lattice sizes, ranging from $8\times 8$, $10\times 10$ to $12\times 12$, using two different approaches. In the first approach,  we introduce and construct the first type of two-dimensional hyperbolic NQS ansatz, the Lorentz 2DRNN, to use in the 2DTFIM setting without mapping to one dimension.
In the second approach, we repeat the 2D-1D mapping experiment (effectively unrolling the 2D setting into a 1D one) of \cite{hld-hypnqs-25} with the new hyperbolic NQS constructions from \cite{hld-hypnqs-26} including Poincar\'e RNN, Lorentz RNN/GRU, in addition to the original Poincar\'e GRU already studied in \cite{hld-hypnqs-25}.  In both cases, we benchmark the performances of all hyperbolic NQS against their Euclidean versions.
\begin{itemize}

\item In native 2D settings with standard (horizontal and vertical) nearest-neighbor interactions — where no structural hierarchy exists — Lorentz 2DRNN consistently outperforms Euclidean 2DRNN at the quantum phase transition point ($B_x = 3.0$), where the system's physics is governed by a conformal field theory (CFT). As a physically motivated heuristic hypothesis, this advantage can be understood through the AdS/CFT correspondence \cite{ads-cft}, where $D$-dimensional CFTs are dual to ($D+1$)-dimensional AdS spaces with hyperbolic spatial geometry.  Consequently, the hyperbolic geometry underlying the construction of the Lorentz 2DRNN naturally provides a well-matched geometric inductive bias for capturing the physics of the TFIM at criticality.

\item In the unrolled 2D settings with one-dimensional NQS, similar to our results in \cite{hld-hypnqs-25},  we find that the one-dimensional hyperbolic NQS models emerge as the better performing ansatzes compared to their Euclidean counterparts thanks to a combination of the induced hierarchical structures determined by the lattice size $N$ and the CFT physics of the system at the critical point.
\end{itemize}
This paper is organized as follows. In Section \ref{sec-phyprnn}, we describe the constructions of the Poincar\'e and Lorentz hyperbolic RNN/GRU, both in one dimension (Section \ref{1d_networks}) and in two dimensions (Section \ref{2d_networks}). In Sec \ref{sec-vmc-summary}, we summarize the variational Monte Carlo (VMC) method, the construction of the neural quantum states based on either Euclidean or hyperbolic RNN/GRU,  as well as some remarks regarding the training of various NQS. The main results of this work are reported in Section \ref{2dtfim_2dnqs_exp} for two-dimensional NQS models and in Section \ref{2dtfim_1dnqs_exp} for one-dimensional NQS models. In Section \ref{concl}, some concluding remarks and future work directions are discussed. In the Appendix, we illustrate the different hierarchical structures representing the different degrees of neighbor interactions in the forms of graphs in Section \ref{graph-forms}. We include the definitions of all mathematical operations in the Poincar\'e disk in Section \ref{ssub:poincar_disk} and in the Lorentz hyperboloid in Section \ref{ssub:lorentz_model} that are used to construct the hyperbolic NQS ansatzes in this work. The full VMC experiment setups and training details of all neural networks are included in Section \ref{ssub:hyperparams}. 
\\\\
The Python codes used in this work to construct the hyperbolic 1D hyperbolic Poincar\'e/Lorentz RNN/GRU  NQS models and 2D hyperbolic Lorentz 2DRNN can be found on our GitHub repository:\\
 \href{https://github.com/lorrespz/hypnqs_tfim}{https://github.com/lorrespz/hypnqs\_tfim}.
 We use a \texttt{Pytorch} implementation to construct all hyperbolic NQS models. 
 The construction of the various Lorentz NQS makes use of the \texttt{hypercore} library \cite{hypercore} with some modifications\footnote{The modifications involve changes in files \texttt{lmath.py} and \texttt{lorentzian.py} from the subdirectory \texttt{manifolds} of the main directory \texttt{hypercore\_main}.}, similar to our previous work \cite{hld-hypnqs-26}.
\section{Poincar\'e and Lorentz hyperbolic RNN/GRU}\label{sec-phyprnn}
In this section, we first recall our constructions of the one-dimensional hyperbolic Poincar\'e and Lorentz RNN/GRU in \cite{hld-hypnqs-25}, \cite{hld-hypnqs-26}, and then proceed to present our custom construction of the two-dimensional hyperbolic Poincar\'e and Lorentz RNN. 
\subsection{One-dimensional networks} \label{1d_networks}
The constructions of the Poincar\'e and Lorentz variants of hyperbolic RNNs were described in detail in our recent work \cite{hld-hypnqs-26}, and are based on the following definitions of Euclidean RNN and GRU: Starting with an input vector $\vec x_i \in \mb{R}^{d_x}$ at step $i$ of size $d_x$, the conventional version of RNN is a function that relates the hidden state vector $\vec h_i\in \mb{R}^{d_h}$ of size $d_h$ to the input $\vec x_i$ and the same hidden state vector $\vec h_{i-1}$ at the previous time step $(i-1)$
\beq
\text{RNN} : && \vec h_i = f(W_h \vec h_{i-1} + U_h\vec x_{i} + \vec b_h)\,,\label{eq-rnn}
\eeq
where $f$ is a nonlinear activation function (often `tanh'), $W_h$ is a $d_h\times d_h$ weight matrix, $U_h$ is a $d_h\times d_x$ weight matrix, and $b_h\in \mb{R}^{d_h}$ is a vector of size $d_h$ known as the bias.
\\\\
GRU \cite{cho-gru} is a more sophisticated version of RNN with additional structures comprising the reset gate $\vec r_i \in \mb R^{d_h}$ and the update gate $\vec z_i\in \mb R^{d_h}$. The defining equations of the conventional or Euclidean GRU are
\beq
\text{GRU}: &&   \vec r_i = \s\lf(W_r   \vec h_{i-1} + U_r \vec x_i +   \vec b_r\rr) \non 
            &&   \vec z_i = \s\lf(W_z   \vec h_{i-1} + U_z \vec x_i +   \vec b_z\rr) \non
            &&   \vec{\tilde h}_i = f\lf[W_h(  \vec r_i \odot  \vec h_{i-1}) + U_h\vec x_i +   \vec b_h\rr] \non
 &&   \vec h_i = (1-\vec z_i)\odot   \vec h_{i-1} +   \vec z_i\odot   \vec{\tilde h}_i \,.\label{gru-update}
\eeq
In Eq.(\ref{gru-update}), $\vec h_i$ is the final hidden state at step $i$ while $\vec{\tilde h}_i$ is the new state computed in the same time step, $f$ is a nonlinear activation function (often taken to be `tanh'), $\s$ is the sigmoid activation, $\vec x_{i}$ is the input vector of length $d_x$, $W_h, W_z, W_r$ are the $d_h\times d_h$ weight matrices, $U_h, U_z, U_r$ are the $d_h\times d_x$ weight matrices, $\vec b_h,   \vec b_z,   \vec b_r \in \mathbb{R}^{d_h}$ are the bias vectors, and $\odot$ is the pointwise multiplication operation.
Note that for GRU, the last equation appearing in Eq.(\ref{gru-update}) that defines the final hidden state $\vec h_i$ can also be written as follows
\beq
\vec h_i = \vec h_{i-1} + \vec z_i\odot(-\vec h_{i-1}+ \vec{\tilde h}_i)\,, \label{eq:gru-update-v2}
\eeq
 where the update step involves the previous state $\vec h_{i-1}$ and the new state $\vec{\tilde h}_i$, which is a function of the reset gate $\vec r_i$, the previous state $\vec h_{i-1}$ and the input $\vec x_i$. When $\vec z_i \approx 0$, the final state $\vec h_i$ is almost entirely $\vec{\tilde h}_i$ - the new state. When $\vec z_i \approx 1$, the final state $\vec h_i$ is almost entirely its previous state $\vec h_{i-1}$. When $\vec r_i \approx 1$ and $\vec z_i \approx 1$, the GRU essentially reduces to the RNN.
\\\\
The Poincar\'e hyperbolic RNN and hyperbolic GRU \cite{hld-hypnqs-25}, \cite{ganea-1805} are defined by the following formulae
\beq
\text{Poincar\'e RNN} : && \vec h_i = f^{\otimes_c}\lf[(W_h \otimes_c \vec h_{i-1}) \oplus_c (U_h\otimes_c \vec x_{i}) \oplus_c \frac{}{}\vec b_h\rr]\, .\label{eq-hrnn}
\\  \non
 \text{Poincar\'e GRU} : && 
  \vec r_i = \s\,\lf\{\log_\mf{0}^c\lf[(W_r\otimes_c \vec h_{i-1}) \oplus_c (U_r\otimes_c \vec x_i) \oplus_c \frac{}{}   \vec b_r\rr]\rr\}\,, \label{hgru-r}
 \non
  && \vec z_i =\s\,\lf\{\log_\mf{0}^c\lf[(W_z\otimes_c \vec h_{i-1}) \oplus_c (U_z\otimes_c \vec x_i) \oplus_c \frac{}{}  \vec b_z\rr]\rr\}\,, \label{eq-hgru}
  \\
  && \vec{\tilde h}_i = f^{\otimes_c}\lf[W_h\,\otimes_c(\vec r_i \odot_c \vec h_{i-1}) \oplus_c (U_h\otimes_c \vec x_i)\frac{}{} \oplus_c   \vec b_h\rr]\,,
  \non
  && \vec h_i =\vec h_{i-1} \oplus_c \,\vec z_i\, \odot_c\lf(-\vec h_{i-1} \oplus_c \vec{\tilde h}_i\rr)\,. \nonumber
 \eeq
In Eqs.(\ref{eq-hrnn}), (\ref{eq-hgru}), the various weight matrices $W_{h,r,z}$ and $U_{h,r,z}$ have the same meanings as in Eqs.(\ref{eq-rnn}), (\ref{gru-update}) for the Euclidean RNN/GRU case, but the hidden state vector $\vec h_i \in \mb{D}^{d_h}_c$, input vector $\vec x_i \in \mb{D}^{d_i}_c$, and bias vectors $\vec b_{h,r,z} \in\mb{D}^{d_h}_c$ are all vectors on the Poincar\'e disk.
Furthermore, unlike the Euclidean case where a nonlinear activation function $f$ (often `tanh') is always applied, we have the choice of either using $f^{\otimes_c}$ or leaving it out completely because hyperbolic space itself provides some degree of nonlinearity. The exact forms of the Poincar\'e mathematical operations $\otimes_c, \oplus_c, \odot_c$, $f^{\otimes_c}$, $\exp_\mathbf{0}^c$, $\log_\mathbf{0}^c$ are given in Section \ref{ssub:poincar_disk}.
\\\\
Other variants of hyperbolic RNN networks, the Lorentz RNN and Lorentz GRU, are based on the Lorentz hyperboloid model of hyperbolic space \cite{hld-hypnqs-26}. Their defining equations, given below, are also based on the Euclidean definitions of RNN and GRU (Eqs.(\ref{eq-rnn}), (\ref{gru-update}))
\beq
\text{Lorentz RNN}: && \vec h_i = f^{\otimes_{\mathcal L}}\lf[\lf(W_h \otimes_\mathcal{L} \vec h_{i-1} \rr) \oplus_\mathcal{L} \lf(U_h \otimes_\mathcal{L} \vec x_i\rr) \oplus_\mathcal{L} \vec b_h\rr] \label{eq-lorentz-rnn} \\
\text{Lorentz GRU}: && \vec r_i = \s\lf\{\log_{\mathbf 0_{\mathcal L}}\lf[\lf(W_r \otimes_\mathcal{L} \vec h_{i-1} \rr) \oplus_\mathcal{L} \lf(U_r \otimes_\mathcal{L} \vec x_i\rr) \oplus_\mathcal{L} \vec b_r\rr] \rr\}\non
&& \vec z_i = \s\lf\{\log_{\mathbf 0_{\mathcal L}}\lf[\lf(W_z \otimes_\mathcal{L} \vec h_{i-1} \rr) \oplus_\mathcal{L} \lf(U_z \otimes_\mathcal{L} \vec x_i\rr) \oplus_\mathcal{L} \vec b_z \rr]\rr\}\label{eq-lorentz-gru}
  \\
  && \vec{\tilde h}_i = f^{\otimes_{\mathcal L}}\lf[W_h\,\otimes_{\mathcal L}(\vec r_i \odot_{\mathcal L} \vec h_{i-1}) \oplus_{\mathcal L} (U_h\otimes_{\mathcal L} \vec x_i)\frac{}{} \oplus_{\mathcal L}  \vec b_h\rr]\,,
  \non
  && \vec h_i = \vec h_{i-1} \oplus_{\mathcal L} \vec z_i\odot_{\mathcal L}\lf(- \vec h_{i-1} \oplus_{\mathcal L} \vec{\tilde h}_{i}\rr)\,, \nonumber
\eeq
where the weight matrices $W_{h,r,z}$, $U_{h,r,z}$ and bias vectors have the same meaning as previously defined. 
\\\\
In Eqs.(\ref{eq-lorentz-rnn}), (\ref{eq-lorentz-gru}), the RNN/GRU hidden state $\vec h_{i}$ at the end of each computation is hyperbolic and lives on the Lorentz hyperboloid.  The weight matrices $W_{h,r,z}, U_{h,r,z}$  are Euclidean, while the biases $\vec b_{h,r,z}$ can be hyperbolic or Euclidean (in which case, they are projected onto the hyperboloid using the exponential map $\exp_{\mathbf{0}_\mathcal L}$). For the Lorentz GRU,  the last equation in Eq.(\ref{eq-lorentz-gru}) is the update step for the hidden state $\vec h_i$, which was done on the hyperboloid in an identical manner to the Poincar\'e case.  However, it must be noted that in Lorentz hyperbolic space, the state $-\vec h_{i-1}$ is not simply the state $\vec h_{i-1}$ with all components multiplied by $-1$ (as is the case in Euclidean space as well as in the Poincar\'e disk). Instead, the $-\vec h_{i-1}$ state has the components  $(x_0, -\vec x_i)$ if $\vec h_{i-1} = (x_0, \vec x_i)$, since we need $\vec h_{i-1} \oplus_{\mathcal L} (-\vec h_{i-1}) = \mathbf 0_{\mathcal L}$.
 The exact forms of the Lorentz mathematical operations $\otimes_\mathcal{L}, \oplus_\mathcal{L}, \odot_\mathcal{L}$, $f^{\otimes_{\mathcal L}}$, $\log_\mathbf{0_\mathcal{L}}$ are described in detail in Section \ref{ssub:lorentz_model}.
\subsection{Two-dimensional networks} \label{2d_networks}
Having recalled the constructions of the 1D hyperbolic NQS in the preceding section, we will now describe the new construction of the two-dimensional hyperbolic Poincar\'e  and Lorentz 2DRNN networks, which are based on the custom construction of the Euclidean 2D RNN described in \cite{rnn_20}:
\beq
\text{Euclidean 2DRNN}: \qquad \vec h_{t+1} = f\lf(U_h \vec x_t^0 + W_h \vec h_t^0 + U_v \vec x_t^1 + W_v \vec h_t^1 + \vec b\rr) \,\label{eq-2drnn}
\eeq
where $f$ is a nonlinear activation function, which can be `tanh' or `elu'. The input $\vec x_t=(\vec x_t^0, \vec x^1_t)$ and RNN hidden state $\vec h_t = (\vec h^0_t, \vec h^1_t)$ at time step $t$ in the computational process are two-dimensional vectors. The number of weight matrices in this case doubles from two to four $W_h, W_v, U_h, U_v$, to take into account the horizontal as well as vertical time steps.
\\\\
The hyperbolic versions of Eq.(\ref{eq-2drnn}) are
\beq
\text{Poincar\'e 2DRNN}: &&   \vec h_{t+1} = f^{\otimes_c}\lf[\lf(U_h \otimes_c \vec x_t^0\rr) \oplus_c \lf(W_h \otimes_c\vec h_t^0\rr) \oplus_c \lf(U_v \otimes_c\vec x_t^1\rr) \oplus_c \lf(W_v\otimes_c \vec h_t^1\rr) \oplus_c \vec b\rr] \\
\label{eq-2drnn-poincare}
\text{Lorentz 2DRNN}: && \vec h_{t+1} = f^{\otimes_\mathcal{L}}\lf[\lf(U_h \otimes_{\mathcal L} \vec x_t^0\rr) \oplus_{\mathcal L} \lf(W_h \otimes_{\mathcal L}\vec h_t^0\rr) \oplus_{\mathcal L} \lf(U_v \otimes_{\mathcal L}\vec x_t^1\rr) \oplus_{\mathcal L} \lf(W_v\otimes_{\mathcal L} \vec h_t^1\rr) \oplus_{\mathcal L} \vec b\rr]
\label{eq-2drnn-lorentz}
\eeq
The Poincar\'e mathematical operations $\otimes_c$, $\oplus_c$, $f^{\otimes_c}$ and the Lorentz mathematical operations $\otimes_{\mathcal L}$, $\oplus_{\mathcal L}$, $f^{\otimes_{\mathcal L}}$ are defined in Section \ref{ssub:poincar_disk} and \ref{ssub:lorentz_model}, respectively.
In this work, we will only focus on the Lorentz 2DRNN due to our limited computing resources, and the fact that in general, Lorentz hyperbolic networks tend to outperform Poincar\'e networks \cite{hld-hypnqs-26}. Furthermore, we note that while it is possible to also construct the hyperbolic Poincar\'e/Lorentz versions of the two-dimensional Euclidean GRU as discussed in \cite{hld-hypnqs-25}, the training of such networks will be very computationally intensive and exceeds our current computational resources.
\newpage
\section{Hyperbolic RNN-based NQS wavefunctions}\label{sec-vmc-summary}
In this section, we summarize the main points regarding the variational Monte-Carlo (VMC) method, as well as the construction of the real NQS wavefunction using either Euclidean or hyperbolic RNN/GRU in discrete Hamiltonian spin systems. The detailed description can be found in \cite{rnn_20}, \cite{hld-hypnqs-25}, \cite{hld-hypnqs-26}.
\\\\
\textbf{Variational Monte Carlo (VMC)}: The VMC method, often used to train NQS, involves the process of sampling from a probability distribution represented by the square of the trial wavefunction/the variational ansatz and subsequently using these generated samples to calculate some observables such as the ground state energy.
In what follows, we recall the derivation of the local energy formula in VMC \cite{mc-textbook}.
\\\\
Given a quantum Hamiltonian $H$, in VMC, the local energy $E_\text{loc}(x)$ of samples $|x\rangle$ generated by an NQS $\Psi$ is given by 
\beq
E_\text{loc}(x) = \sum_{x'} \langle x|H|x'\rangle \frac{\langle x'|\Psi\rangle}{\langle x|\Psi\rangle}. \label{Eloc}
\eeq
$E_\text{loc}(x)$ is non-zero only for those non-zero Hamiltonian elements $\langle x|H|x'\rangle \neq 0$. Corresponding to each generated sample $|x\rangle$ is a probability
\beq
P_\text{loc}(x)= \frac{|\Psi(x)|^2}{\sum_{|x\rangle}|\Psi(x) |^2}\,. \label{Ploc}
\eeq
In the variational problem of interest, given a Hamiltonian $H$ and a trial wavefunction $\Psi$, the VMC task is to estimate the ground state energy $E = \langle \Psi|H|\Psi\rangle$ which can be written in terms of the  local energy $E_\text{loc}$ and the probability $P_\text{loc}(x) $
\beq
E=\sum_{|x\rangle}P_{\text{loc}}(x) E_\text{loc}(x)\,. \label{full-e}
\eeq
When the trial wavefunction is a neural network quantum state $|\Psi(\vec\theta)\rangle$ with trainable parameters $\vec \theta$, at each training iteration $i$, the variational energy $E(\vec\theta)$ is calculated from the local energy $E_\text{loc}(\vec\theta)$ of the Monte Carlo samples generated from the NQS using Eq.(\ref{full-e}). The process of minimizing $E(\vec\theta)$ using an optimizer (such as SGD - Stochastic Gradient Descent or Adam - Adaptive Moment Estimation) updates the trainable parameters $\vec\theta$ until convergence is reached.
\\\\
For the ensuing discussion below regarding the construction of the real NQS ansatzes, the basis state of the Hamiltonian $H$ is denoted by $|\vec\s\rangle = (\s_1, \ldots, \s_N)$ where $N$ is the dimensionality of $H$, and each component $\s_i (1\leq i\leq N)$ assumes discrete values of either 0 or 1.
The real NQS wavefunction is defined as
    \beq
    |\Psi\rangle = \sum_{\vec\s} \sqrt{P(\vec \s)}|\vec \s\rangle
    \eeq
    where $P(\vec\s)$, the probability of a particular configuration $|\vec\s\rangle$, is the output of the real NQS. 
    \beq
    P(\vec \s) = P(\s_1) P(\s_2|\s_1) \ldots P(\s_N|\s_1, \s_2, \ldots, \s_{N-1})
    \eeq

\begin{figure}[H]
\centering
\begin{tikzpicture}[node distance = 1.3cm, thick]%
        \node[circle, draw] (0) {$P(\vec\s_1)$};
        \node[rectangle, draw] (0b) [below of =0, yshift = -0.2cm]{$\begin{array}{c} \text{Dense} \\ (\text{Softmax})\end{array}$};
        \node[rectangle, draw](0c)[below of =0b, yshift=-0.6cm]{$\begin{array}{c}\text{Euclidean/}\\ \text{Hyperbolic}\\ \text{RNN/GRU}\\\end{array}$};
        \node[](0e)[left of = 0c, xshift = -1cm]{};
        \node[circle, draw](0d)[below of =0c, yshift = -0.3cm]{$\s_0$};

        \draw[->] (0d) -- node [right]{} (0c);
        \draw[->] (0c) -- node [right, xshift = 0.1cm]{$h_1$} (0b);
        \draw[->] (0b) -- node [right]{} (0);
        \draw[->] (0e) -- node [right,above, midway]{$h_0$} (0c);

        \node[circle, draw] (1) [right of = 0, xshift=2.2cm]{$P(\vec\s_2)$};
        \node[rectangle, draw] (1b) [below of =1, yshift = -0.2cm]{$\begin{array}{c} \text{Dense} \\ (\text{Softmax})\end{array}$};
        \node[rectangle, draw](1c)[below of =1b, yshift=-0.6cm]{$\begin{array}{c}\text{Euclidean/}\\ \text{Hyperbolic}\\ \text{RNN/GRU}\\\end{array}$};
        \node[circle, draw](1d)[below of =1c,yshift = -0.3cm]{$\s_1$};

         \draw[->] (1d) -- node [right]{} (1c);
        \draw[->] (1c) -- node [right]{$h_2$} (1b);
        \draw[->] (1b) -- node [right]{} (1);
        \draw[->] (0c)-- node [right,above, midway]{$h_1$} (1c);

        \node[circle, draw] (3) [right of = 1, xshift = 2.2cm]{$P(\vec \s_3)$};
        \node[rectangle, draw] (3b) [below of =3, yshift = -0.2cm]{$\begin{array}{c} \text{Dense} \\ (\text{Softmax})\end{array}$};
        \node[rectangle, draw](3c)[below of =3b, yshift=-0.6cm]{$\begin{array}{c}\text{Euclidean/}\\ \text{Hyperbolic}\\ \text{RNN/GRU}\\\end{array}$};
        \node[circle, draw](3d)[below of =3c,yshift = -0.3cm]{$\s_2$};

        \draw[->] (3d) -- node [right]{} (3c);
        \draw[->] (3c) -- node [right]{$h_3$} (3b);
        \draw[->] (3b) -- node [right]{} (3);
        \draw[->] (1c)-- node [right, above, midway]{$h_2$} (3c);

        \node[circle, draw] (4)[above of = 1, yshift= 0.6cm] {$\Psi(\vec\sigma)$};
        \draw[->] (1) -- node [right]{} (4);
        \draw[->] (0) -- node [right]{} (4);
        \draw[->] (3) -- node [right]{} (4);
    \end{tikzpicture}
    \caption{Schematic of the process of calculating the  RNN wavefunction $\Psi(\vec\s)= \sqrt{P(\vec\s)}|\vec\s\rangle$ from the probability $P(\vec\s)$ of the sample $\vec\s$. Here $P(\vec\s) = P(\s_1)P(\s_2|\s_1)\ldots P(\s_N|\s_{N-1})$. For a compact representation, $N=3$ in the schematic. The recurrent network in the diagram can be either Euclidean RNN/GRU or Hyperbolic RNN/GRU. Figure adapted from \cite{hld-hypnqs-25}.} \label{rnn-wavefunc-gen}
    \end{figure}
\FloatBarrier
The autoregressive RNN-based real NQS wavefunction consists of a layer of RNN/GRU followed by a dense layer of two units with the \texttt{Softmax} activation function. The RNN/GRU layer can be either Euclidean or hyperbolic, with the latter choice includes both the options of Poincar\'e and Lorentz models of hyperbolic space. The structure of the RNN-based real NQS function is illustrated in Fig.\ref{rnn-wavefunc-gen}. At a step $i$ where $0\leq i\leq N$, the RNN/GRU cell takes in two arguments, one being the previous RNN/GRU hidden state $\vec h_{i-1}$ and the input $\vec\s_{i-1}$ (the one-hot-encoded $(i-1)^{th}$ component of the generated spin sample $\vec\s=(\s_1, \ldots, \s_N)$), to calculate the current RNN/GRU hidden state $\vec h_i$. This is then passed into the \texttt{Dense} layer with the \texttt{Softmax} activation function\footnote{Recall that the \texttt{Softmax} function is defined as \beq \texttt{Softmax}(v_k) = \frac{\exp(v_k)}{\sum_i\exp(v_i)}\eeq} to obtain an output $\vec y_i$, which is multiplied with the one-hot-encoded sample $\sigma_i$ to calculate the probability $P(\vec\sigma_i)$. The total probability $P(\vec\s)$ for the single spin sample $\vec\s$ is the product of all individual $P(\vec\s_i)$. In this paper, we only work with the real NQS wavefunctions as we are interested in applying the newly constructed 2D hyperbolic NQS Eq.(\ref{eq-2drnn-lorentz}) as well as the 1D hyperbolic NQS Eqs.(\ref{eq-lorentz-rnn}), (\ref{eq-lorentz-gru}) to the 2D TFIM system.
\section{2D TFIM VMC experiments with 2D NQS ansatzes}\label{2dtfim_2dnqs_exp}
The Hamiltonian system of interest to us is the two-dimensional transverse field Ising model (2DTFIM) with open boundary conditions of the form
\beq
H=-J\sum_{\langle i,j\rangle} \s^z_i \s^z_j - B_x\sum_i \s^x_i \label{2dtfim-eq}
\eeq
where $J=1.0$, and the first sum in Eq.\ref{2dtfim-eq} runs over pairs of vertical and horizontal neareast neighbors.
This system exhibits a phase transition at $B_c = 3.044$, separating an ordered, ferromagnetic phase with $\langle \s^z\rangle \neq 0$ from a disordered, paramagnetic phase where $\langle \s^z\rangle = 0$ \cite{2dtfim-mc}. At $B_x \approx B_c$ near the critical magnetic field strength $B_c$, the 2DTFIM is described by a  3D conformal field theory (3DCFT)\footnote{Recall that 2DTFIM at criticality where $B_x=B_c$ is equivalent to a 3D classical Ising model at criticality where $T=T_c$ \cite{sachdev}, which can be described by a 3D conformal field theory \cite{cft-book}.
For the correspondence between $D$-dimensional classical Ising model at criticality and ($D+1$)-dimensional CFT, refer to Chapter 12 of the book \cite{cft-book}.} \cite{cft-book}.  At the critical point, the system, described the 3DCFT physics, exhibits a long range, power-law spin spin correlation $\langle \sigma^z_0 \sigma_r^z\rangle$ of the form\footnote{Note that away from the critical point, this spin-spin correlation has the exponential decay form $\langle \sigma_0^z \sigma_r^z\rangle \sim e^{-r/\xi}$, where $\xi$ is a characteristic length scale depending on the system.}
\beq
\langle \sigma_0^z \sigma_r^z\rangle = \frac{A}{r^{D-2+\eta}} \,,\label{spin-spin}
\eeq
where $D$ is the number of spatial dimensions, $\eta$ is a constant that varies depending on the  number of spatial dimensions, $A$ is a proportionality constant chosen depending on the system under study. For $D=3$ (corresponding to 3D classical Ising model dual to 2D TFIM), $\eta=0.0363$ \cite{3d-ising-cft}. The critical exponent $\gamma= D-2+\eta$ is a universal quantity that defines the phase transition, but the proportionality constant $A$ is a non-universal prefactor that depends on the microscopic cutoff and operator normalization of specific discrete lattice simulations, so $A$ can be treated as a free scaling parameter to vertically align the profiles of the spin-spin correlation curves in systems under study.
\\\\
 In this work, we will carry out different sets of VMC experiments using different types of NQS ansatzes. These experiments are designed to benchmark the performances of our newly constructed hyperbolic NQS (both one-dimensional and two-dimensional) ansatzes against the performances of the established Euclidean RNN ansatzes. The detailed description of all experiment settings and parameters can be found in Section \ref{ssub:hyperparams}. 
\\\\
 In this section, we report the results of the VMC experiments in which we apply the newly constructed hyperbolic Lorentz 2DRNN NQS in  Eq.(\ref{eq-2drnn-lorentz}) in the 2DTFIM setting on a square lattice\footnote{In this work, we use $N\times N$ and ($N,N)$ interchangeably to denote the square lattice size of the 2DTFIM.}. There are two subsets of experiments, one involving different lattice sizes $(N,N)=(8,8), (10,10), (12,12)$ at fixed magnetic field strength $B_x=3.0$ (almost at criticality), and one involving the same lattice size $(N,N) = (12,12)$ at different magnetic field strengths $B_x = 2.0, 3.0, 4.0$. For each $B_x$ and each lattice size $N$, a VMC experiment involving one NQS ansatz is done with three independent NQS random weight initialization seeds to ensure variability across independent optimization runs. We note that while a larger seed ensemble is desirable, for $N=12$, hyperbolic NQS optimization trajectories often require up to 20 hours or more of compute time, making higher seed counts computationally prohibitive under our current resource budget (for more details, refer to Section \ref{ssub:hyperparams}). 
\\\\
For the 2DTFIM experiments with 2D ansatzes, the NQS models and their corresponding parameters are listed in Table \ref{fig-2d-ansatz} (see Table \ref{tfim2d-hyperparams} in Section \ref{ssub:hyperparams} for all VMC experiment parameters). For $(N,N)=(8,8)$, the RNN hidden vector size of all ansatzes is 50, while for $(N,N)=(10,10)$, (12,12), the hidden vector size is 60. For Lorentz 2DRNN, in some experiments, we consider several different ansatzes corresponding to different choices of the spatial constraint hyperparameter $L_{max}$\footnote{$L_{max}$ is a hyperparameter, introduced in \cite{hld-hypnqs-26}, that caps the spatial components of all Lorentz vectors in the Lorentz NQS to ensure numerical stability of the ansatz during training. Different choices of $L_{max}$ lead to different levels of expressivity, hence different Lorentz ansatzes. See Section \ref{ssub:hyperparams} for full details.}.
\begin{table}[!ht]
\centering
\begin{tabular}{cccccc}
\hline\hline
& Ansatz & Hidden dimension size & Parameters & Experiment &\\
\hline\hline
&Euclidean 2DRNN & 50 & 5352 &  $(N,N) = (8,8)$ & \\
& & 60 & 7622 &  $\begin{array}{c} (N,N) = (10,10)\\(N,N) = (12,12) \end{array}$ & \\
\hline
&Lorentz 2DRNN & 50 & 5352  &  $(N,N) = (8,8)$ & \\
& & 60 & 7622 &  $\begin{array}{c} (N,N) = (10,10)\\(N,N) = (12,12) \end{array}$ & \\
\hline\hline
\end{tabular}
\caption{The two-dimensional NQS ansatzes used to run the 2DTFIM VMC experiments involving different lattice sizes ($N,N$). For $(N,N)=(8,8)$, the NQS ansatzes have a fixed RNN hidden dimension size of 50, while for (10,10) and (12,12), the RNN hidden dimension size is 60.} \label{fig-2d-ansatz}
\end{table}

\FloatBarrier
\subsection{Different lattice sizes, fixed magnetic strength} \label{sub:varied_N_fixed_B}
\begin{figure}[!ht]
\centering
\includegraphics[width=.7\textwidth]{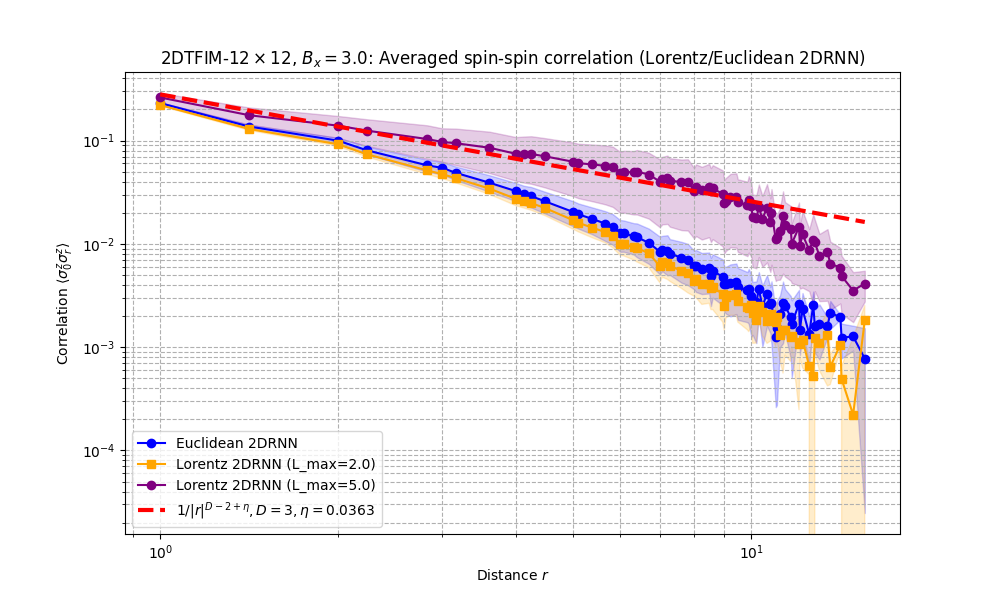}
\caption{The spin-spin correlation decay length curves versus distance $r$ for Lorentz 2DRNN (with $L_{max}=2.0$ and $L_{max}=5.0$) and Euclidean 2D RNN NQS ansatzes  averaged across three different NQS seeds in 2DTFIM VMC experiments with the lattice size of $(N,N)$= (12,12) at a fixed magnetic field strength $B_x=3.0$. The shaded region around each curve corresponds to the standard error obtained from averaging over the NQS seeds. The VMC sampling uncertainty errors are negligible by comparison in magnitude and too small to be shown. The red dashed line corresponds to the 3D CFT power-law spin-spin correlation  $\langle \s^z_0\s^z_r\rangle  \propto 1/|r|^{D-2+\eta}$ where $D=3$ and $\eta=0.0363$. For the 2D lattice setting, the distance between two spins $(i_1,j_1)$ and $(i_2,j_2)$ on the 2D lattice is calculated as $r=\sqrt{(i_1-i_2)^2 + (j_1-j_2)^2}$. We restricted the $r$ range to $10^1$ to filter out the fluctuations due to finite size effects and open boundary conditions. }\label{fig-spin-corr-N12-ave}
\end{figure}

\begin{figure}[!ht]
\centering
\includegraphics[width=.7\textwidth]{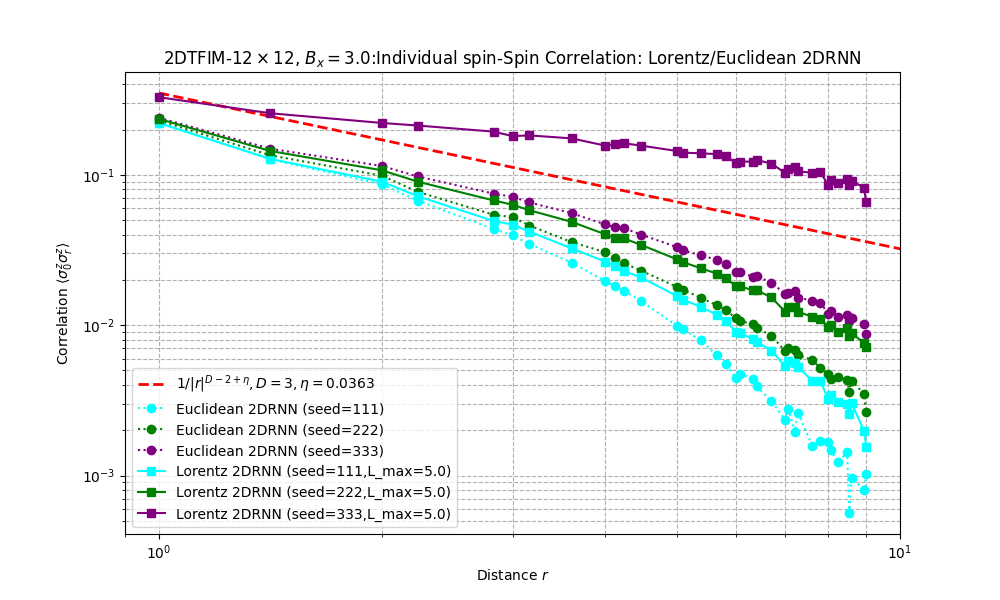}
\caption{The individual spin-spin correlation decay length curves for Lorentz 2DRNN and Euclidean 2D RNN NQS ansatzes in 2DTFIM VMC experiments with the lattice size of $(N,N)$= (12,12) at a fixed magnetic field strength $B_x=3.0$. The error bars due to VMC sampling uncertainties are too small to be shown. The red dashed line corresponds to the 3D CFT power-law spin-spin correlation $\langle \s^z_0\s^z_r\rangle  \propto 1/|r|^{D-2+\eta}$ where $D=3$ and $\eta=0.0363$. When seed=111, the two curves are identical and overlap. For the 2D lattice setting, the distance between two spins $(i_1,j_1)$ and $(i_2,j_2)$ on the 2D lattice is calculated as $r=\sqrt{(i_1-i_2)^2 + (j_1-j_2)^2}$.  We restricted the $r$ range to $10^1$ to filter out the fluctuations due to finite size effects and open boundary conditions. Note that while seed=333 displays a significantly flatter correlation profile than the other initializations, its final converged energy remains highly accurate and close to the true ground state.}\label{fig-spin-corr-N12-ind}
\end{figure}

\begin{figure}[!ht]
\centering
\includegraphics[width=.7\textwidth]{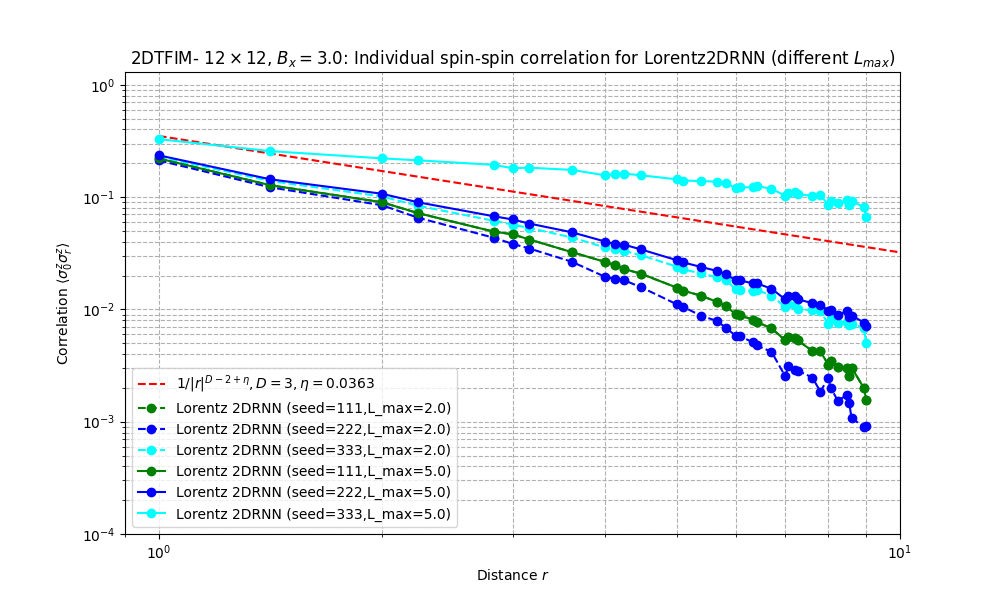}
\caption{The individual spin-spin correlation decay length curves for Lorentz 2DRNN ansatzes  (with $L_{max}=2.0$ and $L_{max}=5.0$) in 2DTFIM VMC experiments with the lattice size of $(N,N)$= (12,12) at a fixed magnetic field strength $B_x=3.0$. The error bars due to VMC sampling uncertainties are too small to be shown. The red dashed line corresponds to the 3D CFT power-law spin-spin correlation $\langle \s^z_0\s^z_r \rangle \propto 1/|r|^{D-2+\eta}$ where $D=3$ and $\eta=0.0363$. For the 2D lattice setting, the distance between two spins  $(i_1,j_1)$ and $(i_2,j_2)$ on the 2D lattice is calculated as $r=\sqrt{(i_1-i_2)^2 + (j_1-j_2)^2}$. We restricted the $r$ range to $10^1$ to filter out the fluctuations due to finite size effects and open boundary conditions. }\label{fig-spin-corr-N12-ind-2}
\end{figure}

The results of the VMC experiments with different ($N,N$) at fixed $B_x=3.0$ are shown in Tables \ref{2dtfim_varied_N8_res}, \ref{2dtfim_varied_N10_res}, \ref{2dtfim_varied_N12_res} and Fig.\ref{fig-B-res}. In these tables and figures, we list both the best and averaged achievable mean energies, where `best' refers to the best result chosen from all different individual VMC runs, while `averaged' refers to the averaged result taken from all VMC runs with different NQS random weight initialization seeds.
 In Fig.\ref{fig-spin-corr-N12-ave} and Fig.\ref{fig-spin-corr-N12-ind}, we also plotted the $\langle \s^z_0\s^z_r\rangle$ spin-spin correlation length decay $\langle \s^z_0\s^z_r\rangle$ for all 2D NQS ansatzes for the case of $(N,N)=(12,12)$ using the averaged results from different VMC runs and the individual results from each run, respectively\footnote{Unlike the unrolled case considered in the next section where we have a 2D to 1D mapping in which each lattice size $N$ represents a different hierarchical neighbor interaction structure $\langle i,i+1\rangle$-$\langle i, i+N\rangle$, in the natively two-dimensional setting, only nearest neighbor interactions (both horizontal and vertical) are present, and the `hierarchical' structure (or the lack thereof) is the same for all lattice sizes. It suffices, therefore, to only consider the spin-spin correlation curves of all 2D NQS ansatzes for the largest lattice size of $N=12$.}.
\begin{table}[!ht]
\centering
\begin{tabular}{c cc c}
\hline\hline
 & $\begin{array}{c} \text{NQS Ansatz} \end{array}$ &  $\begin{array}{cc}(N,N)=(8,8),  & B_x=3.0\\ \text{Best}  &\text{Averaged}  \end{array}$ &\\
\hline\hline
& Euclidean 2DRNN &
$\begin{array}{cccc} & -202.4230 & -202.3914& \\ &  0.0099&  0.0108&\\\end{array}$ & \\\\
& Lorentz 2DRNN ($L_{max}=2.0)$&
$\begin{array}{cccc} & \mathbf{-202.4597}&  \mathbf{-202.4538} &\\ &0.0074 & 0.0076  & \\\end{array}$  & \\\\
\hline
& DMRG  & -202.5077  &\\\hline
\end{tabular}
\caption{VMC results (both best and averaged from different runs with distinct random NQS weight initialization seeds) of 2D TFIM with the lattice size $(N,N)$ = (8,8) at the fixed magnetic field strength $B_x=3.0$. For each entry, we record the mean energy in the upper line, and the associated standard error (corresponding to VMC sampling) in the lower line. The best-performing results are noted in bold. } \label{2dtfim_varied_N8_res}
\end{table}
\begin{table}[!ht]
\centering
\begin{tabular}{c cc c}
\hline\hline
  & $\begin{array}{c} \text{NQS Ansatz} \end{array}$ &  $\begin{array}{cc}(N,N)=(10,10),  & B_x=3.0\\ \text{Best}  &\text{Averaged}  \end{array}$ &\\
\hline\hline
& Euclidean 2DRNN &
$\begin{array}{cccc} & -316.8991 & -316.8717& \\ &  0.0089&  0.0103&\\\end{array}$ & \\\\
& Lorentz 2DRNN ($L_{max}=2.0)$&
$\begin{array}{cccc} & \mathbf{-316.9353} & \mathbf{-316.8919} &\\ &0.0077 & 0.0099 & \\\end{array}$  & \\\\
\hline
& DMRG   & -316.9770 &
\\\hline
\end{tabular}
\caption{VMC results (both best and averaged from different runs with distinct random NQS weight initialization seeds) of 2D TFIM with varied size lattice $(N,N) =$ (10,10) at the fixed magnetic field strength $B_x=3.0$. For each entry, we record the mean energy in the upper line, and the associated standard error (corresponding to VMC sampling) in the lower line. The best-performing results are noted in bold. } \label{2dtfim_varied_N10_res}
\end{table}
\begin{table}[!ht]
\centering
\begin{tabular}{c cc c}
\hline\hline
 & $\begin{array}{c} \text{NQS Ansatz} \end{array}$ &  $\begin{array}{cc}(N,N)=(12,12),  & B_x=3.0\\ \text{Best}  &\text{Averaged}  \end{array}$ &\\
\hline\hline
& Euclidean 2DRNN &
$\begin{array}{cccc} & -456.8887 &  -456.8643 & \\ &   0.0122&  0.0123&\\\end{array}$ & \\\\
& Lorentz 2DRNN ($L_{max}=2.0)$&
$\begin{array}{cccc} & -456.9595 &   \textbf{-456.9121} &\\ & 0.0084&0.0104& \\\end{array}$  & \\\\
& Lorentz 2DRNN ($L_{max}=5.0)$ &
$\begin{array}{cccc} &\textbf{-456.9670}& -456.8606 &\\ & 0.0088& 0.0124
 &\end{array}$ & \\\\
\hline
& DMRG & -457.0416  &\\
\hline
\end{tabular}
\caption{VMC results (best and averaged from different runs with distinct random NQS weight initialization seeds) of 2D TFIM with the lattice size $(N,N)$ = (12,12) at the fixed magnetic field strength $B_x=3.0$. For each entry, we record the mean energy in the upper line, and the associated standard error (corresponding to VMC sampling) in the lower line. The best-performing results are noted in bold. } \label{2dtfim_varied_N12_res}
\end{table}
\FloatBarrier
From the results recorded in  Tables \ref{2dtfim_varied_N8_res} - \ref{2dtfim_varied_N12_res}, the following observations are made.
\begin{itemize}
  \item $(N,N)$ = (8,8), (10,10):
  \bei\item
   Lorentz 2DRNN (with $L_{max}=2.0$) robustly outperforms Euclidean 2DRNN for both lattice sizes, both in terms of the best and averaged mean energy reachable.
   \eni
  \item $(N,N)$ = (12,12):
  \bei \item
  In this case, we consider two different Lorentz 2DRNN NQS ansatzes, one with $L_{max}=2.0$ and one with $L_{max}=5.0$. In terms of the best energy reachable from multiple individual runs, Lorentz 2DRNN with $L_{max}=5.0$ is the best ansatz, but in terms of the averaged energy reachable from all individual runs taken together, Lorentz 2DRNN with $L_{max}=2.0$ is the best ansatz.
  \item In terms of the averaged spin-spin correlation decay length  $\langle \s^z_0\s^z_r\rangle $ (see the log-log plots in Fig.\ref{fig-spin-corr-N12-ave}), Lorentz 2DRNN with $L_{max}=5.0$ shows the best performance, in the form of an averaged curve that conforms best to the 3D CFT line\footnote{On a log-log scale, the power-law relation Eq.(\ref{spin-spin}) becomes a straight line and the exponent $\gamma=D-2+\eta$ in the denominator becomes the slope of this line.} compared to both Euclidean 2DRNN and Lorentz 2DRNN with $L_{max}=2.0$. The latter two show very close spin-spin correlation decay curves, with Euclidean 2DRNN performing slightly better than Lorentz 2DRNN with $L_{max}=2.0$.  In Fig.\ref{fig-spin-corr-N12-ind}, where individual correlation curves resulting from individual VMC runs are shown, the Lorentz 2DRNN ($L_{max}=5.0$) curve shows a slower correlation decay, with a slope that better matches the power-law exponent in Eq.(\ref{spin-spin}) than  the corresponding Euclidean 2DRNN's curve for every single random seed considered. In Fig.\ref{fig-spin-corr-N12-ind-2}, when comparing the $L_{max}=2.0$ against the $L_{max}=5.0$ for each VMC run (using the same seed), the curves belonging to the $L_{max}=5.0$ curve lie well above those belonging to $L_{max}=2.0$ for two out of three runs (in the remaining run, the two curves are identical).
  \item While both Lorentz 2DRNN considered in this case could outperform the Euclidean 2DRNN in terms of lower reachable energy, their behaviors are very different when it comes to capturing the long range correlation  near the phase transition point, with one Lorentz 2DRNN clearly performing much better than the other. This highlights the importance of choosing the right $L_{max}$ to ensure the optimal performance of the hyperbolic Lorentz 2DRNN.
  \eni

  \item Overall, in this series of experiments with fixed magnetic field strength $B_x=3.0$ at criticality, Lorentz 2DRNN has shown a consistent and robust outperformance compared to Euclidean 2DRNN. This is interesting because this probably has very little to do with the structure of the neighbor interactions, which is not that hierarchical (since it only comprises nearest neighbor interactions) compared to the unrolled 2D case studied in \cite{hld-hypnqs-25} and also in the next section. The natural question that arises is, if not because of the hierarchical advantage associated with hyperbolic space that has led to the outperformance of hyperbolic NQS compared to their Euclidean counterpart as seen previously in \cite{hld-hypnqs-25}, \cite{hld-hypnqs-26}, what could account for this observed performance trend ?
\\\\
While a formal mathematical proof exceeds the scope of this work, to answer this question (speculatively), we propose a physically motivated heuristic hypothesis, which starts with the fact that at criticality, the physics of a $(D-1)$-dimensional quantum TFIM system can be described by $D$-dimensional CFT \cite{cft-book}, \cite{sachdev}. CFTs in $D$ spatial dimensions, on the other hand, are known from the celebrated AdS/CFT correspondence \cite{ads-cft}, to be dual to an Anti-de-Sitter (AdS) space in $(D+1)$ dimensions, whose spatial geometry is that of a $D$-dimensional hyperbolic space.
 In this particular case under study, we have 2DTFIM at criticality described by a 3DCFT dual to an $AdS_4$ space with a hyperbolic $\mathbb H^3$ spatial geometry. In this sense, the two-dimensional hyperbolic space underlying the construction of the Lorentz 2DRNN \footnote{Despite the apparent 2D-3D mismatch in the dimensionality of the hyperbolic space $\mathbb H^2$ in which the Lorentz 2DRNN is constructed and the hyperbolic space $\mathbb H^3$ that is the spatial section of the $AdS_4$ space dual to a 3DCFT, hyperbolic Lorentz 2DRNN still possesses the `correct' geometry to represent the ground state wavefunction of the 2DTFIM at criticality much better than Euclidean 2DRNN. } might be the likely reason why hyperbolic Lorentz 2DRNN NQS outperforms its Euclidean counterpart at a point where the system displays conformal field theory behaviors\footnote{Recall that in the 1DTFIM  VMC experiments performed at the critical field strength $B_x=1.0$ in \cite{hld-hypnqs-25}, for $N=80$ spins, 1D Poincar\'e GRU NQS did not outperform Euclidean GRU NQS but for $N=100$ spins Poincar\'e GRU NQS did. However, these results are not complete in the sense that other types of 1D hyperbolic NQS ansatzes such as Lorentz RNN/GRU  and Poincar\'e RNN were not included at that time since the experiments were performed prior to their proposed constructions in \cite{hld-hypnqs-26}. Also, the construction of the Poincar\'e GRU NQS in \cite{hld-hypnqs-25} was restricted to the case where the nonlinear activation $f^{\otimes_c}$ in Eq.\ref{eq-hgru} is the identity function while other possibilities like `tanh' and `elu' exist.}.
 \\\\
Other supporting evidences for the connection between the underlying hyperbolic geometry of Lorentz 2DRNN NQS playing a decisive role in its outperformance over Euclidean 2DRNN NQS in TFIM at criticality come from the seminal works \cite{vidal-05}, \cite{vidal-06} that proposes a type of variationial ansatz known as MERA (Multi-scale Entanglement Renormalization Ansatz) that is shown to be particularly suited to represent quantum ground states at criticality \cite{vidal-06}. In particular, MERA is a type of tensor network ansatz consisting of two types of tensors - isometries ($w$) and disentanglers ($u$) - arranged in a hierarchical, tree-like manner. Its winning point over other types of tensor network ansatzes \cite{tn-18}, \cite{tn-13}, such as DMRG in 1D and PEPS (Projected Entangled Pair States) \cite{peps-0}, \cite{peps-1}, \cite{peps-2} or TTN (Tree Tensor Network) in 2D lies in the fact that it was specifically designed to represent quantum states at criticality with long range entanglement and power-law correlation \cite{vidal-06}, \cite{vidal-08}. In \cite{swingle-09}, it is realized that the discrete minimal-cut paths through a MERA network is mathematically identical to the spatial geodesic in a continuous 2D hyperbolic/AdS geometry. This means that MERA is a discrete, skeletal version of an emergent holographic spacetime, where the vertical layer depth ($z$) is the extra radial dimension of AdS space \cite{swingle-12}.
 \\\\
 In other words, given that MERA, the well-established tensor ansatz designed to simulate critical quantum states, is fundamentally discrete hyperbolic geometry, it stands to reason that our hyperbolic NQS constructions with their natively continuous hyperbolic geometry are able to automatically exploit the natural exponential volume expansion of the underlying space to capture the power-law correlations of TFIM critical quantums states much more efficiently than Euclidean NQS ever could. Furthermore, while MERA runs into the problem of increasing computational complexity in 2D because of the exponential number of tensors required in their construction, 1D and 2D hyperbolic NQS ansatzes constructed in this work and in our previous works \cite{hld-hypnqs-25}, \cite{hld-hypnqs-26} do not have this problem since they leverage continuous neural network optimization while capturing the same hyperbolic geometry advantage.
\end{itemize}
\begin{figure}[!ht]
\centering
\includegraphics[width=.6\textwidth]{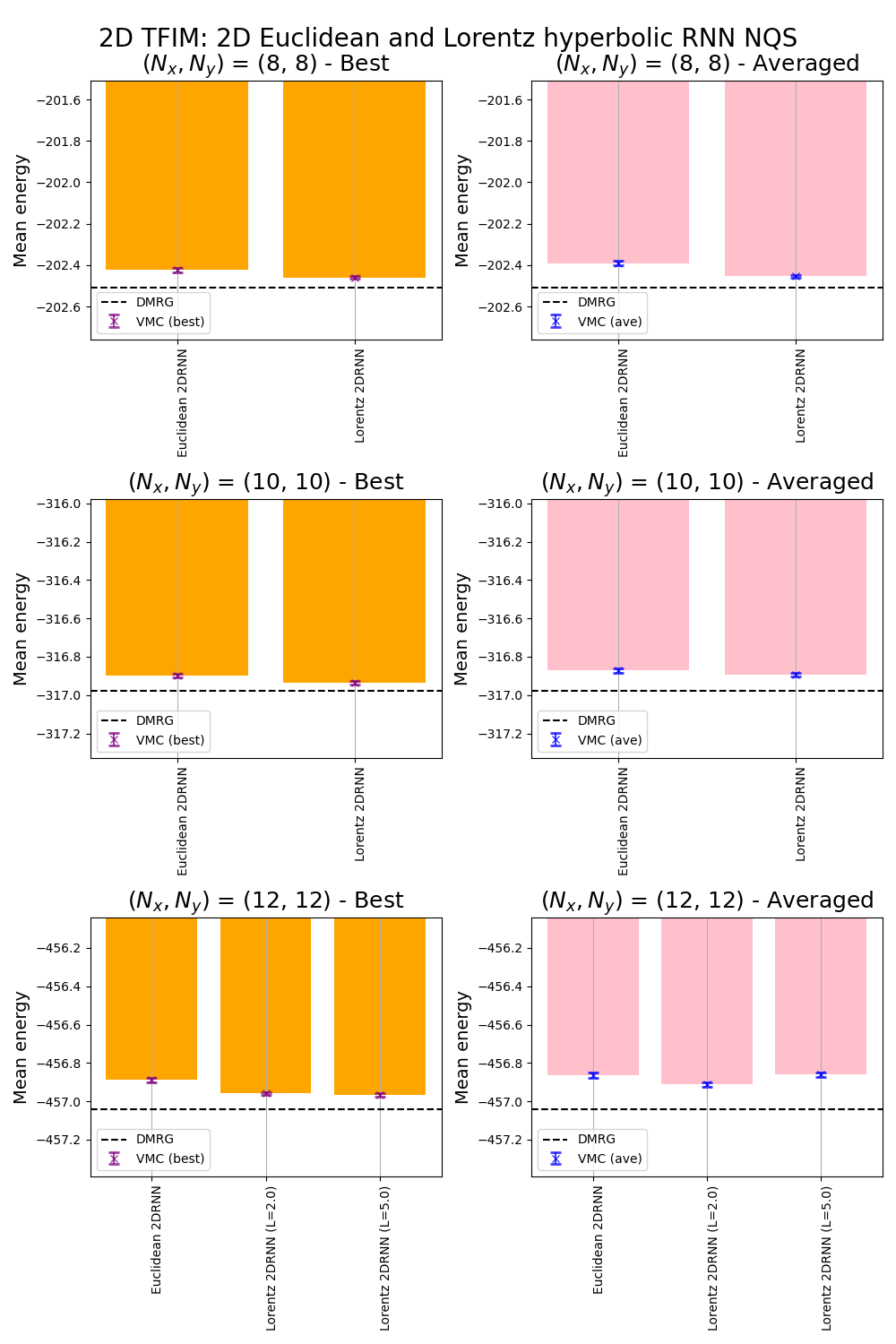}
\caption{A comparison of the performances of Euclidean 2DRNN and Lorentz 2DRNN  (both best and averaged from different runs with distinct random NQS weight initialization seeds) in terms of the mean ground state energy in the VMC experiments involving 2DTFIM with the lattice sizes of $(N,N)$=(8,8), (10,10), (12,12) at a fixed magnetic field strength $B_x=3.0$. The error bars in all plots correspond to VMC sampling noise. For $(8,8)$ and (10,10) cases, the Lorentz 2DRNN has $L_{max}=2.0$ while for $(12,12)$, two different Lorentz 2DRNN NQS, with $L_{max}=2.0, 5.0$ were considered.}\label{fig-B-res}
\end{figure}
\FloatBarrier
\FloatBarrier

\FloatBarrier

\FloatBarrier
\newpage
\subsection{Fixed lattice sizes, different magnetic strengths}
While it is shown above that Lorentz 2DRNN outperform Euclidean 2DRNN quite robustly at criticality when $B_x=3.0$, we are also interested in the performances of Lorentz 2DRNN when it comes to different magnetic field strengths away from the critical point. As such, we fix the lattice size of $(N,N)$=(12,12) and look at the cases where $B_x=2.0, 3.0, 4.0$. The results of these experiments (both best and averaged) are recorded in Table \ref{2dtfim_varied_B2} for $B_x=2.0$ and Table \ref{2dtfim_varied_B4} for $B_x=4.0$. For $B_x=3.0$, the results were already listed in Table \ref{2dtfim_varied_N12_res} in the previous section.
\begin{table}[!ht]
\centering
\begin{tabular}{c cc c}
\hline\hline
 & $\begin{array}{c} \text{NQS Ansatz} \end{array}$ &  $\begin{array}{cc} (N,N)=(12,12), & B_x = 2.0 \\\text{Best}  &\text{Averaged}  \end{array}$ &\\
\hline\hline
& Euclidean 2DRNN &
$\begin{array}{cccc} & \mathbf{-346.8500} &  \mathbf{-346.7618}  & \\ & 0.0087&  0.0111 &\\\end{array}$ & \\\\
& Lorentz 2DRNN ($L_{max}=2.0)$&
$\begin{array}{cccc} & -346.3433 &   -346.3065 &\\ & 0.0183&0.0188& \\\end{array}$  & \\\\
& Lorentz 2DRNN ($L_{max}=6.0)$ &
$\begin{array}{cccc} & -346.4035  & -346.3426 &\\ & 0.0172&0.0185
 &\end{array}$ & \\\\
\hline
& DMRG  & -346.9828  &\\
\hline
\end{tabular}
\caption{VMC results (best and averaged from different runs with distinct random NQS weight initialization seeds) of 2D TFIM with the fixed size lattice $(N,N)$ = (12,12) at the transverse magnetic field strength $B_x=2.0$. For each entry, we record the mean energy in the upper line, and the associated standard error (corresponding to VMC sampling) in the lower line.  The best-performing results are noted in bold. } \label{2dtfim_varied_B2}
\end{table}
\begin{table}[!ht]
\centering
\begin{tabular}{c cc c}
\hline\hline
 & $\begin{array}{c} \text{NQS Ansatz}\ \end{array}$ &  $\begin{array}{cc} (N,N)=(12,12), & B_x = 4.0\\  \text{Best}  &\text{Averaged}  \end{array}$ &\\
\hline\hline
& Euclidean 2DRNN &
$\begin{array}{cccc} & \mathbf{-593.5006} & -593.3540    & \\ & 0.0072&  0.0144 &\\\end{array}$ & \\\\
& Lorentz 2DRNN ($L_{max}=2.0)$&
$\begin{array}{cccc} &-593.4953   & -593.3216  &\\ &0.0077 &0.0139 & \\\end{array}$  & \\\\
& Lorentz 2DRNN ($L_{max}=1.5)$ &
$\begin{array}{cccc} & -593.4719& \mathbf{-593.4648} &\\ & 0.0068  &0.0079
 &\end{array}$ & \\\\
\hline
& DMRG  & -593.5389  &\\
\hline
\end{tabular}
\caption{VMC results (best and averaged from different runs with distinct random NQS weight initialization seeds) of 2D TFIM  with the fixed size lattice $(N,N) =$ = (12,12) at the transverse magnetic field strength $B_x=4.0$. For each entry, we record the mean energy in the upper line, and the associated standard error (correponding to VMC sampling) in the lower line.  The best-performing results are noted in bold. } \label{2dtfim_varied_B4}
\end{table}
\FloatBarrier
From Tables \ref{2dtfim_varied_N12_res}, \ref{2dtfim_varied_B2},  \ref{2dtfim_varied_B4}, the following observations are made.
\bei
\item  At $B_x=3.0$ (Table \ref{2dtfim_varied_N12_res}): When  $B_x=3.0$, the system is at the phase transition point with long-range spin-spin correlation Eq.(\ref{spin-spin}) and Lorentz 2DRNN robustly outperform Euclidean 2DRNN. As discussed in detail in the previous section, this observed trend could be accounted for by a physically motivated hypothesis involving the fact that the physics of the system at this point displays conformal symmetry and describable by a CFT, which is dual to an AdS space whose spatial geometry is hyperbolic. Subsequently, compared to the flat geometry of Euclidean space, the hyperbolic geometry underlying the construction of Lorentz 2DRNN likely provides a better match to represent the ground state wavefunction of the system at criticality.
\item At $B_x=2.0$ (Table \ref{2dtfim_varied_B2}):  When $B_x=2.0$, the ground state is an ordered ferromagnetic system, and Euclidean 2DRNN outperform Lorentz 2DRNN (with $L_{max}=2.0$ and $L_{max}=6.0$) robustly, both in terms of best and averaged reachable ground state energy. In this case, where the entire ground state is ordered and uniform, the flat Euclidean geometry wins over the hyperbolic one, since there is no hierarchical structure nor CFT physics that offers hyperbolic geometry an advantage.

\item  At $B_x=4.0$ (Table \ref{2dtfim_varied_B4}): When  $B_x=4.0$, the system is a disordered paramagnet, and interestingly, the performances of Euclidean 2DRNN and Lorentz 2DRNN are comparable. In terms of the absolute best energy obtained from individual VMC runs, Euclidean 2DRNN emerges as the better NQS, while in terms of the averaged best energy obtained from all VMC runs taken together, Lorentz 2DRNN (with $L_{max}=1.5$) emerges as the more stable and better-performing NQS.

\eni
\begin{figure}[!ht]
\centering
\includegraphics[width=.7\textwidth]{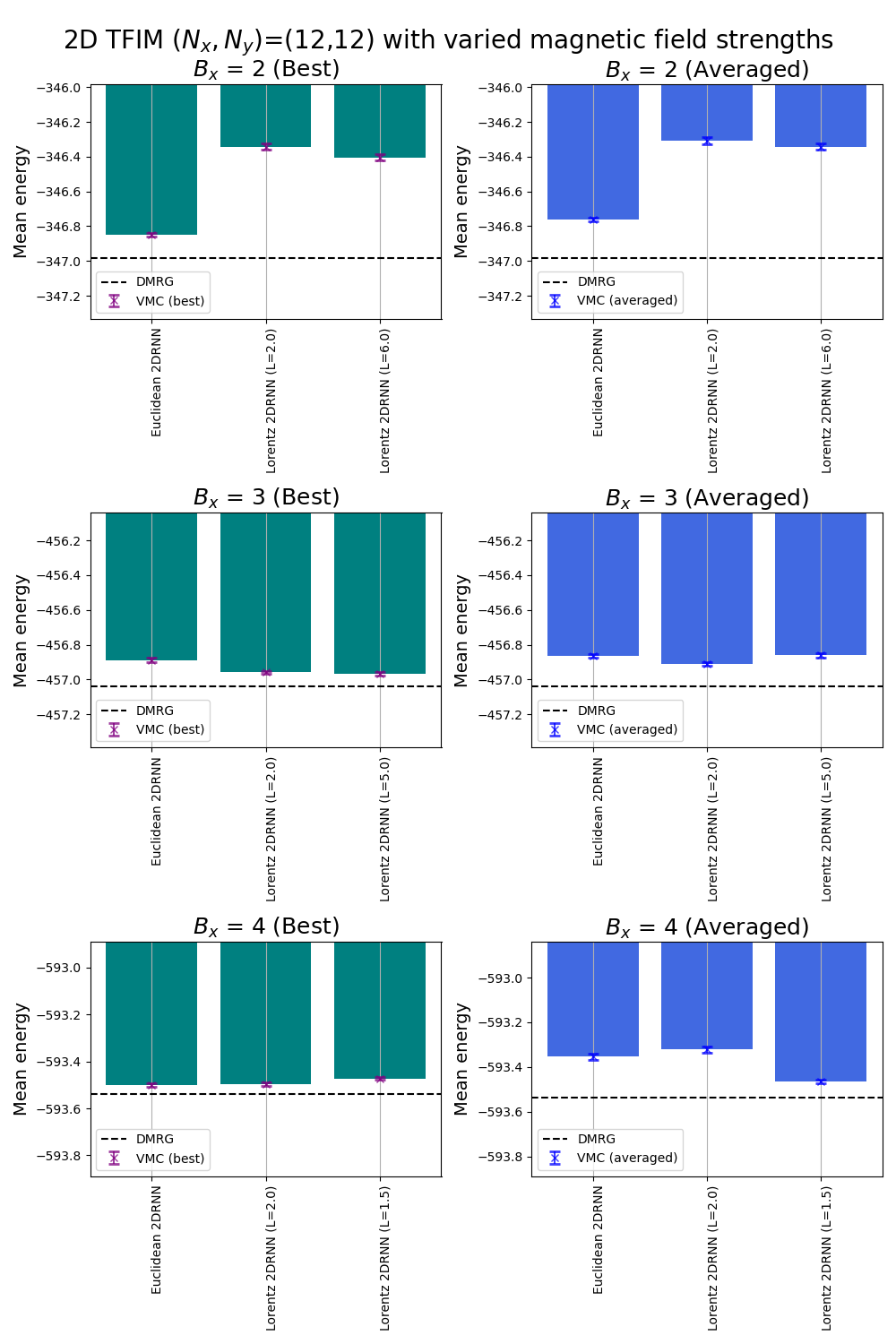}
\caption{A comparison  of the performances (both best and averaged from different VMC runs with distinct random NQS initialization seeds) of Euclidean 2DRNN and Lorentz 2DRNN models in terms of the mean ground state energy in the VMC experiments involving 2DTFIM with the lattice size of $(N,N)=(12,12)$ at three different magnetic field strengths $B_x=2.0, 3.0, 4.0$. The error bars in all plots correspond to VMC sampling noise.}\label{fig-B-res}
\end{figure}
\FloatBarrier
\newpage
\section{2D TFIM VMC experiments with 1D NQS ansatzes} \label{2dtfim_1dnqs_exp}
\begin{figure}[!ht]
\centering
\includegraphics[width=.6\textwidth]{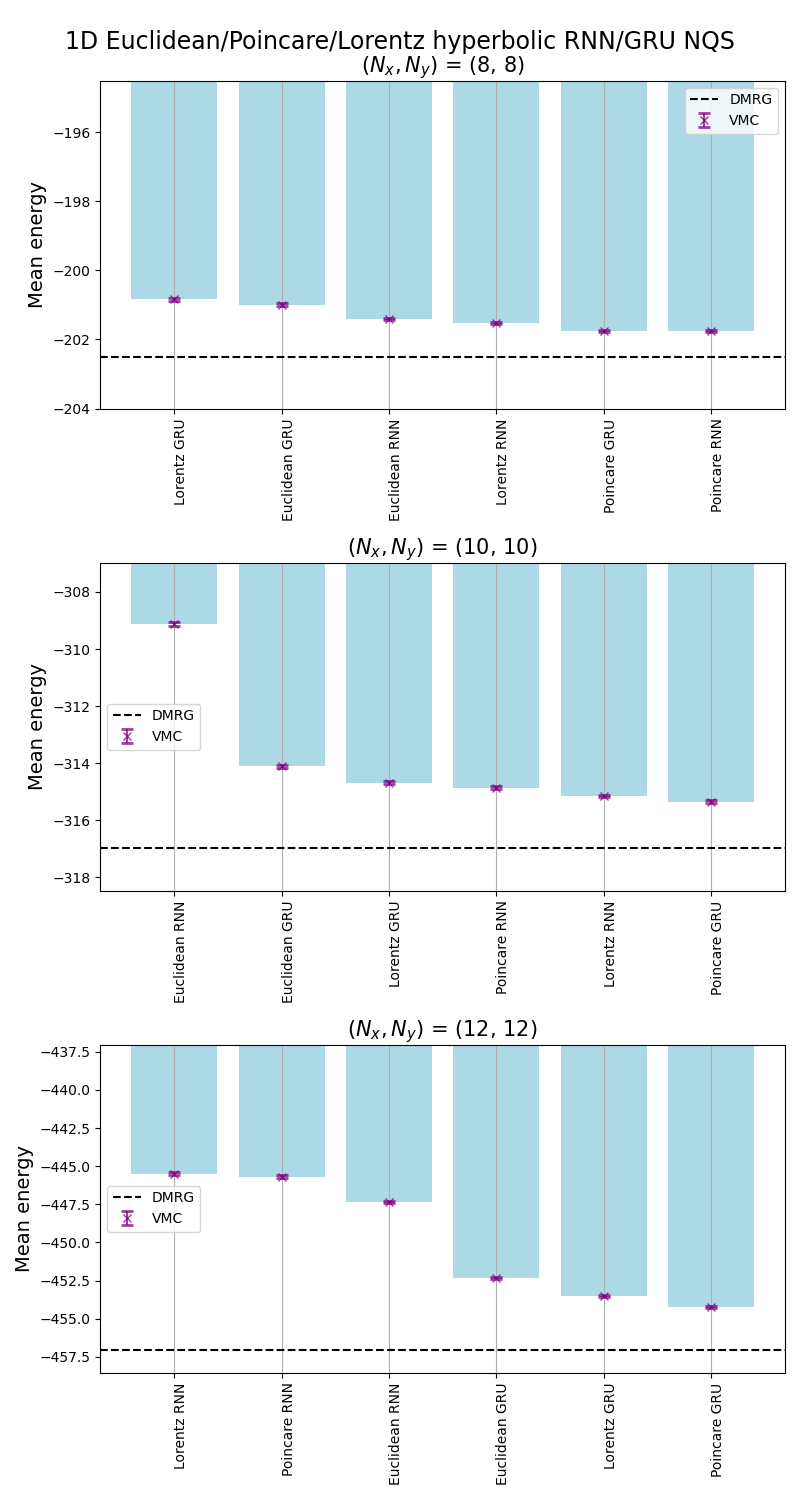}
\caption{The best performances of 1D Euclidean and hyperbolic Poincar\'e/Lorentz NQS ansatzes, chosen from different runs corresponding to different NQS random initialization seeds, ranked in ascending order for each of the 2D TFIM VMC experiment setting $(N,N) = (8,8)$, (10,10), (12,12). The error bars correspond to uncertainties in the VMC sampling process. } \label{1dnqs_vmc_ranking}
\end{figure}

\begin{figure}[!ht]
\centering
\includegraphics[width=.6\textwidth]{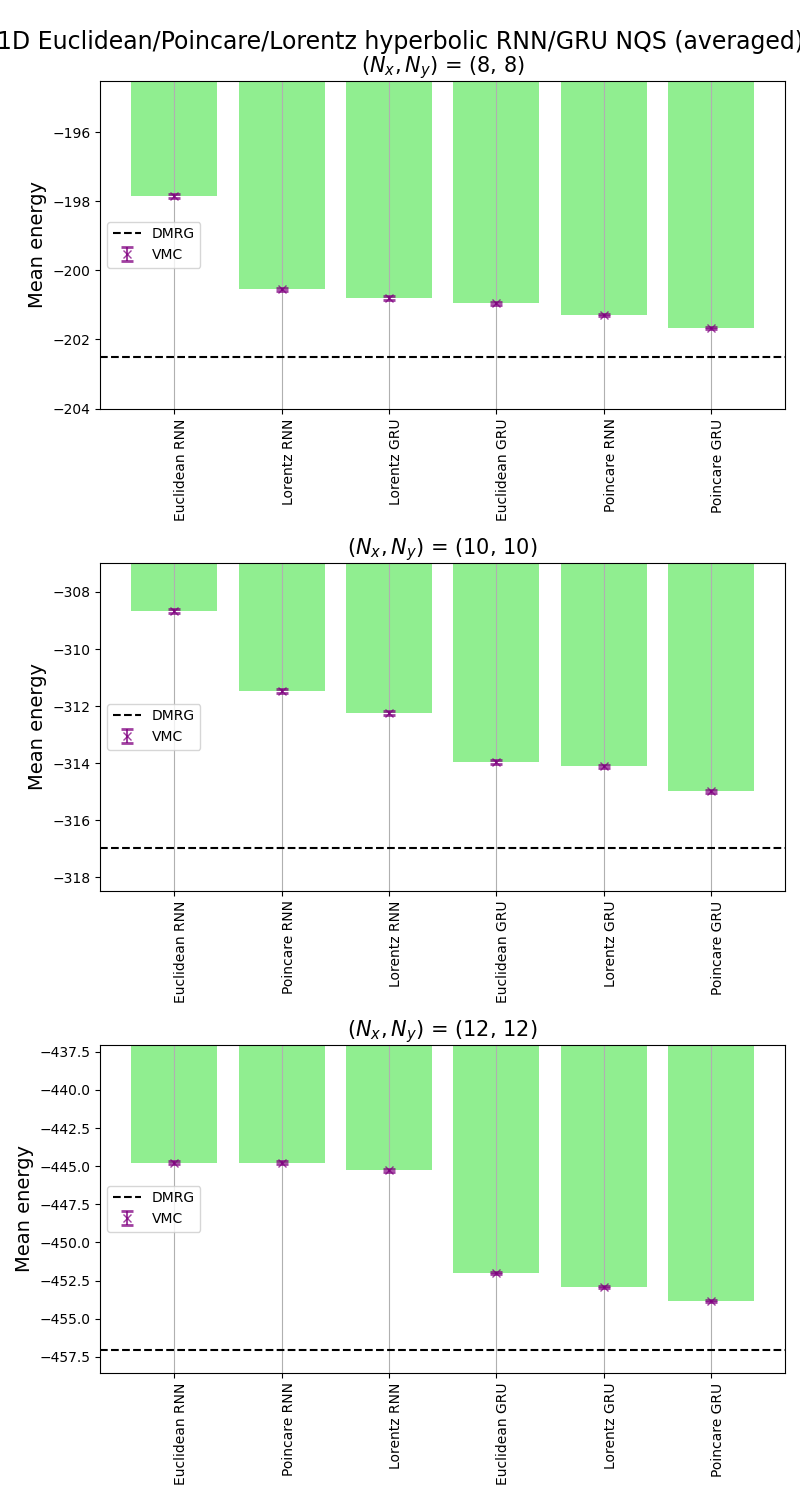}
\caption{The performances of 1D Euclidean and hyperbolic Poincar\'e/Lorentz NQS ansatzes, averaged over different runs corresponding to different NQS random initialization seeds, ranked in ascending order for each of the 2D TFIM VMC experiment setting $(N,N) = (8,8)$, (10,10), (12,12). The error bars correspond to uncertainties in the VMC sampling process.  } \label{1dnqs_vmc_ranking_averaged}
\end{figure}

\begin{figure}[!ht]
\centering
\includegraphics[width=.8\textwidth]{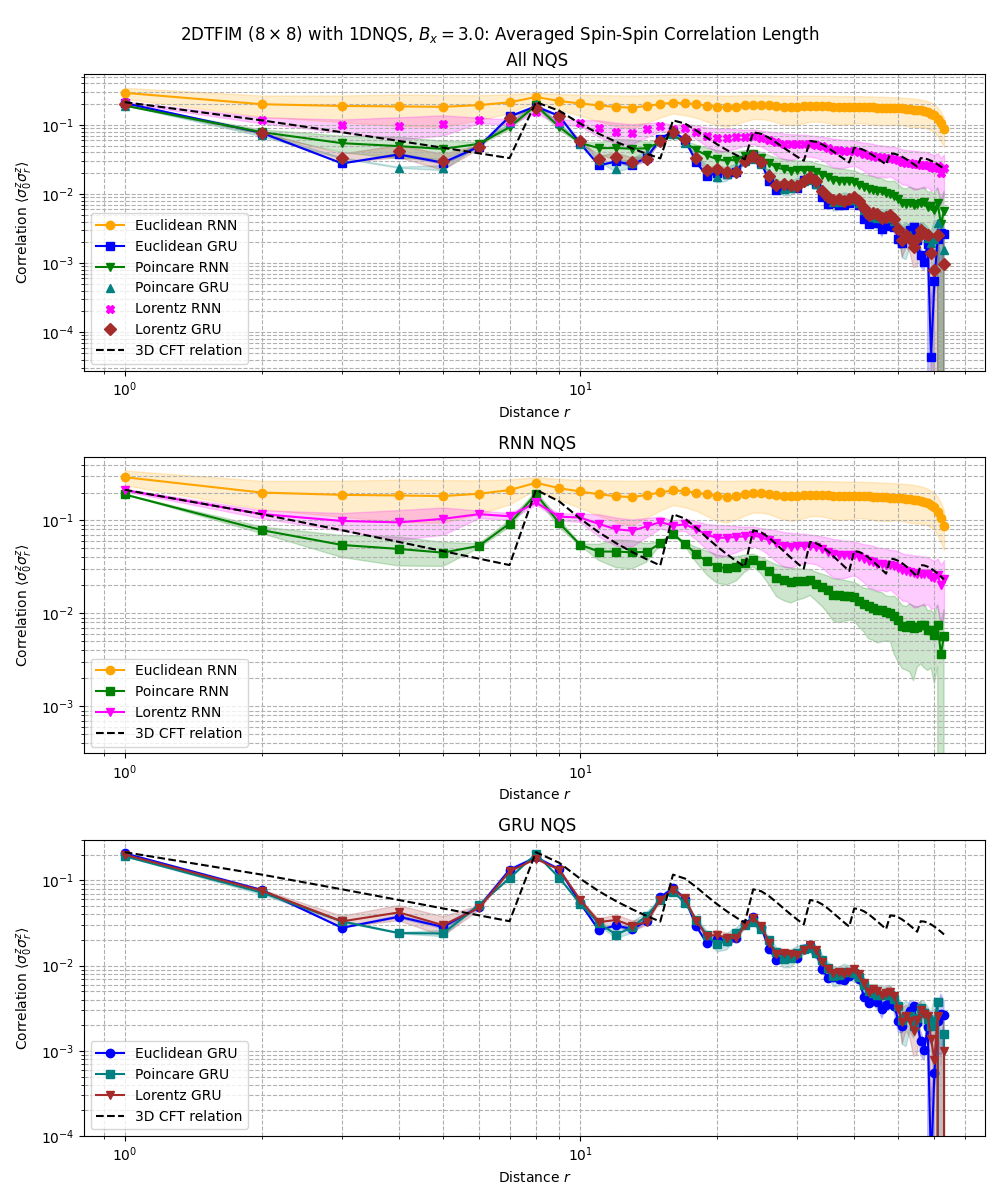}
\caption{The $\langle \sigma^z_0\sigma_r^z\rangle$ spin-spin correlation decay length curves for all six types of 1D NQS ansatzes from the unrolled 2D TFIM experiments with $(N,N) = (8,8)$. The shaded region around each curve corresponds to the standard error obtained from averaging over the NQS random initialization seeds. The VMC sampling uncertainties are negligible by comparison and too small to be shown. From top to bottom: In the top subfigure, the curves for all six ansatzes are shown together, in the middle subfigure, the zoomed-in curves of the three RNN-based NQS variants are shown, in the bottom subfigure, the zoomed-in curves of the three GRU-based NQS variants are shown. Note that due to the open boundary conditions, the correlation length curves should decay much faster at larger distances (due to the decrease of neighbors for spin sites nearing the lattice edge). The fluctuations seen at large $r$ are due to finite size effects of the lattice. The black, dashed line is the 3D CFT power relation Eq.(\ref{spin-spin}) translated to one-dimensional setting as in Eq.\ref{3d-cft-1d}. The exact 3D CFT baseline exhibits a sharp sawtooth profile because it maps a continuous, isotropic physical power-law ($r^{-d}$) onto a discrete 1D unrolled sequence of 2D lattice coordinates.  } \label{ss-corr-N8-1dnqs}
\end{figure}

\begin{figure}[!ht]
\centering
\includegraphics[width=.8\textwidth]{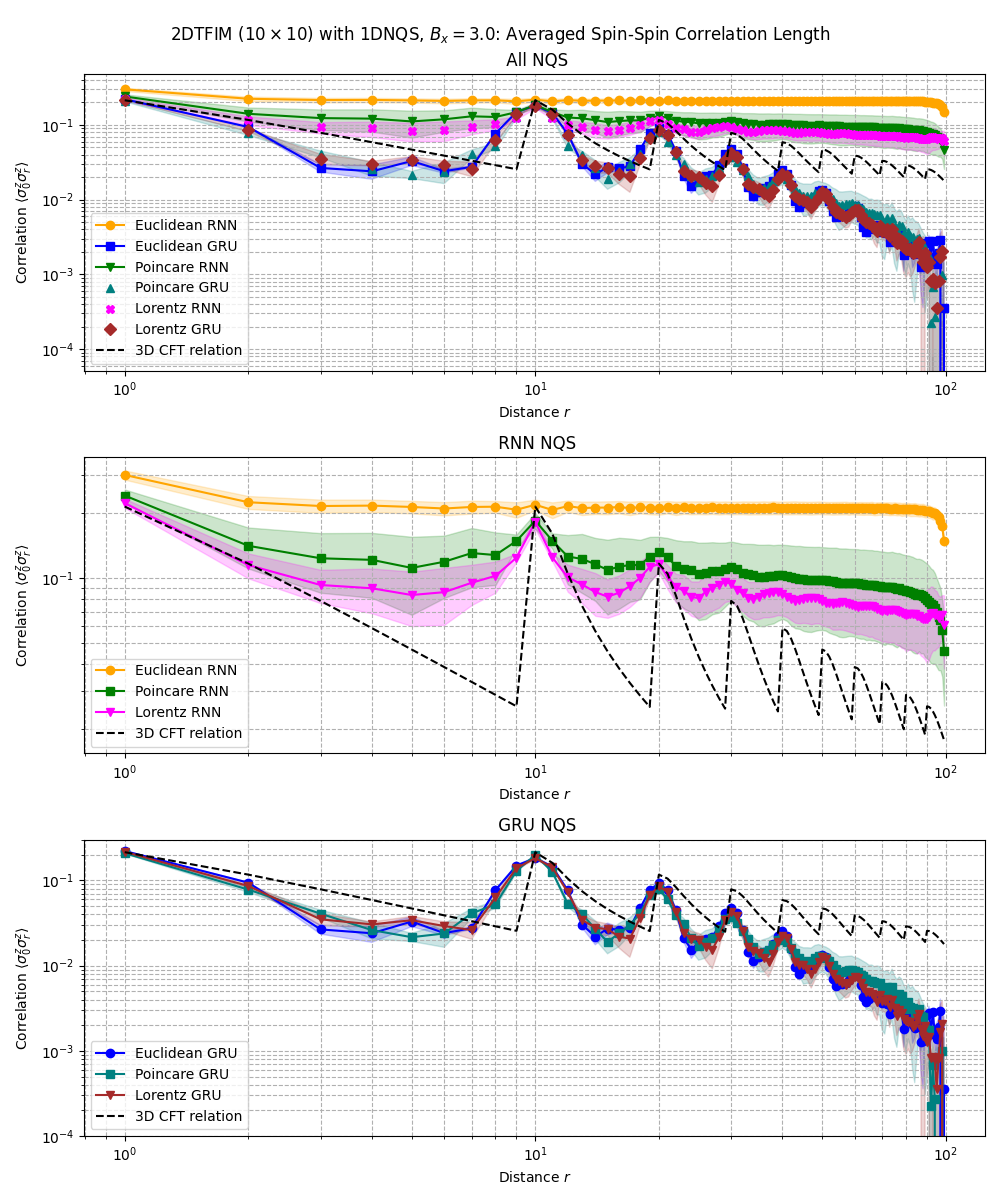}
\caption{The $\langle \sigma^z_0\sigma_r^z\rangle$ spin-spin correlation decay length curves for all six types of 1D NQS ansatzes from the unrolled 2D TFIM experiments with $(N,N) = (10,10)$. The shaded region around each curve corresponds to the standard error obtained from averaging over the NQS random initialization seeds. The VMC sampling uncertainties are negligible by comparison and too small to be shown. From top to bottom: In the top subfigure, the curves for all six ansatzes are shown together, in the middle subfigure, the zoomed-in curves of the three RNN-based NQS variants are shown, in the bottom subfigure, the zoomed-in curves of the three GRU-based NQS variants are shown. Note that due to the open boundary conditions, the correlation length curves should decay much faster at larger distances (due to the decrease of neighbors nearing the lattice edge). The fluctuations seen at large $r$ are due to finite size effects of the lattice. The black, dashed line is the 3D CFT power relation Eq.(\ref{spin-spin}) translated to one-dimensional setting as in Eq.\ref{3d-cft-1d}. The exact 3D CFT baseline exhibits a sharp sawtooth profile because it maps a continuous, isotropic physical power-law ($r^{-d}$) onto a discrete 1D unrolled sequence of 2D lattice coordinates.} \label{ss-corr-N10-1dnqs}
\end{figure}

\begin{figure}[!ht]
\centering
\includegraphics[width=.8\textwidth]{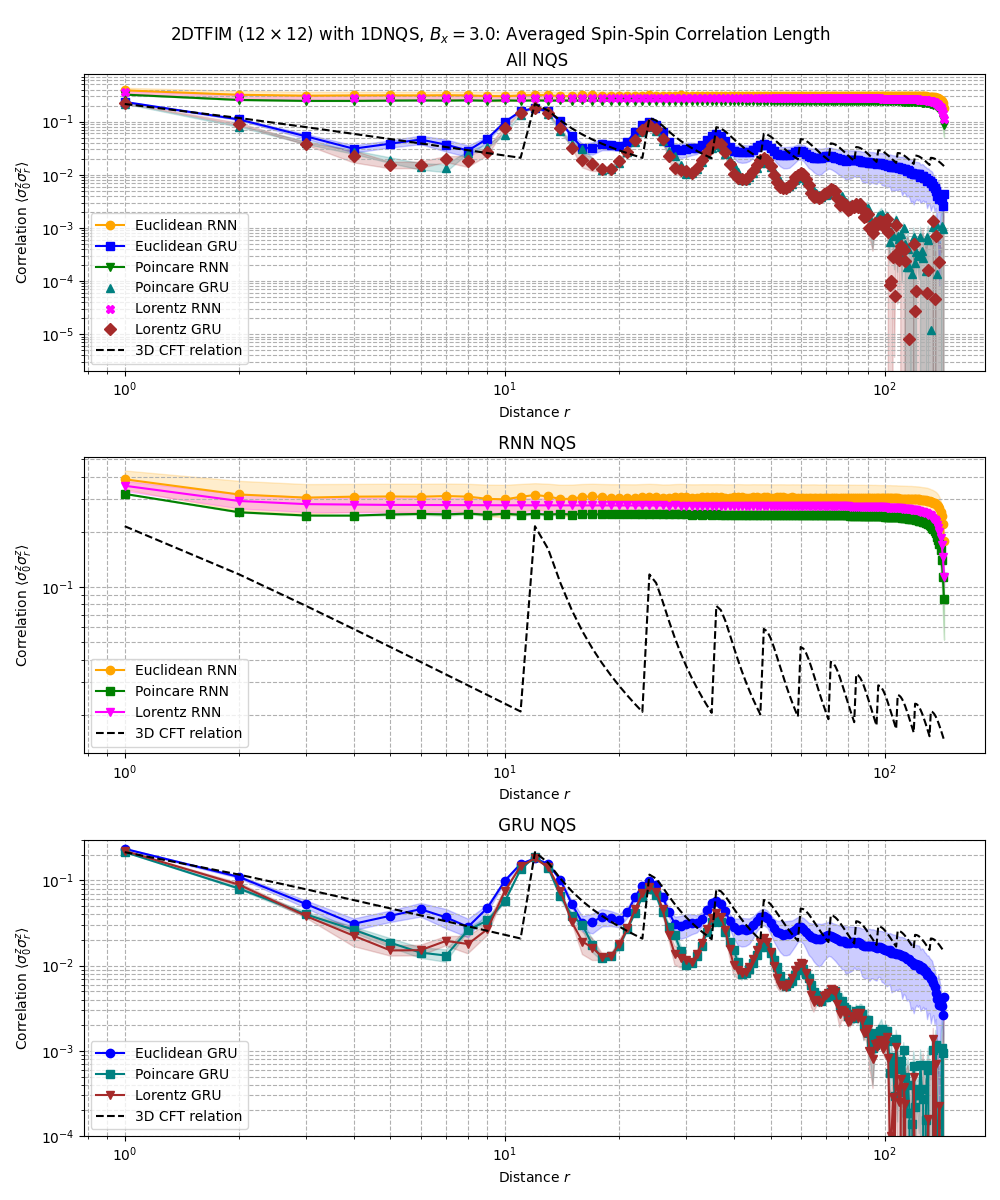}
\caption{The $\langle \sigma^z_0\sigma_r^z\rangle$ spin-spin correlation decay length curves for all six types of 1D NQS ansatzes from the unrolled 2D TFIM experiments with $(N,N) = (12,12)$. The shaded region around each curve corresponds to the standard error obtained from averaging over the NQS random initialization seeds. The VMC sampling uncertainties are negligible by comparison and too small to be shown. From top to bottom: In the top subfigure, the curves for all six ansatzes are shown together, in the middle subfigure, the zoomed-in curves of the three RNN-based NQS variants are shown, in the bottom subfigure, the zoomed-in curves of the three GRU-based NQS variants are shown. Note that due to the open boundary conditions, the correlation length curves should decay much faster at larger distances (due to the decrease of neighbors nearing the lattice edge). The fluctuations seen at large $r$ are due to finite size effects of the lattice. The black, dashed line is the 3D CFT power relation Eq.(\ref{spin-spin}) translated to one-dimensional setting as in Eq.\ref{3d-cft-1d}. The exact 3D CFT baseline exhibits a sharp sawtooth profile because it maps a continuous, isotropic physical power-law ($r^{-d}$) onto a discrete 1D unrolled sequence of 2D lattice coordinates.} \label{ss-corr-N12-1dnqs}
\end{figure}

In this section, we report the results of the VMC experiments involving the recently constructed one-dimensional hyperbolic Poincar\'e  and Lorentz RNN/GRU NQS ansatzes in \cite{hld-hypnqs-26} when these are applied to the problem of two-dimensional TFIM in the same manner as done in \cite{hld-hypnqs-25}. Our main aim in carrying out this type of experiments is to understand the representational capacity or expressivity of hyperbolic NQS ansatzes versus their Euclidean counterparts in capturing the hierarchical neighbor interaction structure in the 2DTFIM setting unrolled in one dimension. As described in detail in \cite{hld-hypnqs-25}, in translating or unrolling from two dimensions to one, the nearest horizontal and vertical neighbor interactions of the 2D $N\times N$ square lattice become the $(i,i+1)$ and $(i, i+N)$ interactions in one dimension.
\\\\
Specifically, if we start with a two-dimensional $N\times N$ square lattice of spins whose sites are labeled ($i_{2D}, j_{2D}$) with $1\leq i_{2D}, j_{2D} \leq N$, in two dimensions, the horizontal and vertical nearest neighbor pairs are
\beq
\lf\langle (i_{2D}, j_{2D}), (i_{2D}+1, j_{2D})\rr\rangle, \qquad \lf\langle (i_{2D}, j_{2D}), (i_{2D},j_{2D} +1 )\rr\rangle.
\eeq
When treated as a one-dimensional system, this 2D square lattice becomes a spin chain of length $N^2 = (N-1)N + N$, with the following mapping of site location:
\beq
\lf( i_{2D}, j_{2D}\rr) \rightarrow \lf[(i_{2D}-1)N + j_{2D} \rr] \label{2d-1d-map}
\eeq
which leads to the 2D nearest neighbor interactions becoming the following 1D neighbor interactions, upon substituting $i_{1D} = (i_{2D}-N) + j_{2D}$\footnote{For later convenience, we also included the reverse mapping from 1D spin chain to 2D lattice: \beq i_{2D} = (i_{1D}//N) +1 \qquad j_{2D}=(i_{1D} \,\text{mod}\, N) +1 \,.\label{2d-1d-mapping} \eeq }:
\beq
\lf\langle (i_{2D}, j_{2D}), (i_{2D}+1, j_{2D}) \rr\rangle & \rightarrow & \lf\langle i_{1D}, i_{1D}+1\rr\rangle  \non
\lf\langle  (i_{2D}, j_{2D}), (i_{2D},j_{2D} +1 )\rr\rangle & \rightarrow & \lf\langle i_{1D}, i_{1D} + N\rr\rangle
\eeq
This means that the mapping of 2DTFIM into the setting of 1D TFIM results in the original 2D nearest neighbor interactions becoming the first neighbor interaction $\langle i, i+1\rangle$ and  $N^{th}$ neighbor interactions $\langle i, i+N\rangle$ in one dimension. Thus, different 2D lattice sizes $N$ result in inherently different 1D hierarchical structures\footnote{In Figs.\ref{graph-like-h} and \ref{graph-like-t} in the Appendix, we illustrate different neighbor interaction hierarchical structures of different spin systems in the form of graphs.}, in constrast to keeping the 2D setting intact where different 2D lattice size $N$ does not change the nature of the interaction being fundamentally first-neighbor interactions only (along horizontal and vertical dimensions).
In \cite{hld-hypnqs-25}, we hypothesized that this hierarchy of interactions between the first and $N^{th}$ neighbor interactions was the main reason why 1D hyperbolic NQS (in the form of Poincar\'e GRU) outperformed its Euclidean version, the Euclidean GRU NQS in 2DTFIM systems of up to $(N,N) = (9,9)$. In this work, we expand this result to larger systems including $(N,N) = (10,10), (12,12)$, with additional types of 1D hyperbolic NQS ansatzes including Poincar\'e RNN, Lorentz RNN and Lorentz GRU, in addition to the originally introduced Poincar\'e GRU in \cite{hld-hypnqs-25}.
\\\\
The six types of 1D NQS used to run the 2DTFIM VMC experiments are listed in Table \ref{1dnqs_detail}, while Table \ref{tfim2d-ur-hyperparams} in Section \ref{ssub:hyperparams} lists all VMC experiment parameters. Similar to \cite{hld-hypnqs-26}, these six NQS types encompass two architecture variants: RNN and GRU, with two underlying geometries: Euclidean and hyperbolic (which further divided into the subgeometry of Poincar\'e disk and Lorentz hyperboloid). For these 1D NQS ansatzes, we are only interested in the experiments with varying lattice sizes at the fixed magnetic field strength of $B_x=3.0$, which is practically at the phase transition point (since $B_c=3.044$).  In Table \ref{1dnqs_vmc_results}, we list the best VMC results corresponding to each of these NQS (where the best is chosen from different VMC runs of one particular NQS ansatz), while in Table \ref{1dnqs_vmc_results_ave}, we list the averaged results obtained from all different VMC runs (corresponding to different NQS random initialization seeds). In Fig.\ref{1dnqs_vmc_ranking}, we show their best performances in ascending order, while in Fig.\ref{1dnqs_vmc_ranking_averaged}, we show their performances averaged over different VMC runs in ascending order.
\\\\
Identical to the 2D NQS case studied above, in addition to computing the mean ground state energy that each 1D NQS ansatz is capable of reaching, we also compute the spin-spin correlation length decay $\langle \sigma^z_0\sigma^z_r\rangle$ of each of the six 1D NQS ansatzes for each lattice size $N=8,10,12$ (shown in Figs.\ref{ss-corr-N8-1dnqs}, \ref{ss-corr-N10-1dnqs}, \ref{ss-corr-N12-1dnqs}). It is important to note that for this particular situation where we have a one-dimensional setting arising from artificially unrolling the two dimensional system using Eq.(\ref{2d-1d-map}), a `good' 1D NQS is not one that is capable of producing a straight-line decaying curve conforming to Eq.(\ref{spin-spin}), but one that is capable of learning the inherent two-dimensional geometry of the lattice. In other words, a good 1D NQS ansatz in this case would be one that is capable of capturing the $N^{th}$-neighbor interactions $\langle i, i+N\rangle$, and this shows up in the spin-spin correlation curve as a periodic pattern with the period corresponding exactly to the 2D lattice size $N$. In fact, the power-law correlation Eq.(\ref{spin-spin}), from the 3D CFT that is equivalent to the 2DTFIM at criticality, when translated to the unrolled 1D setting, is
\beq \langle \s^z_o\s^z_r\rangle_{1D} = \frac{A}{\left[ i_{2D}^2 + j_{2D}^2 \right]^{1.0363/2}},\label{3d-cft-1d}
\eeq
where $A$ is a proportionality constant chosen depending on the system under study, $i_{2D}$ and $j_{2D}$ are the 2D lattice coordinates written in terms of 1D spin chain site $i_{1D}$ as defined in Eq.(\ref{2d-1d-mapping}).
Eq.(\ref{3d-cft-1d}) describes an oscillating, periodic pattern whose period exactly coincides with the 2D lattice size $N$. More specifically, Eq.(\ref{3d-cft-1d}) is not a smooth sine-wave-like curve but rather a sawtooth waveform because of the discreteness of the spin site locations.
In Figs. \ref{ss-corr-N8-1dnqs}, \ref{ss-corr-N10-1dnqs}, \ref{ss-corr-N12-1dnqs}, the 3DCFT power relation in 1D form Eq.(\ref{3d-cft-1d}) is also included as a reference to benchmark the performances of different NQS ansatzes in terms of their abilities to capture the correct spin-spin correlation decay characteristics.
\begin{table}[!ht]
\centering
\begin{tabular}{clcccc}
\hline\hline
& Ansatz & Hidden dimension size & Parameters & Experiment &\\
\hline\hline
&Euclidean RNN & 60 &  3902 &  $(N,N) = (8,8)$ & \\
& & 70 & 5252 &  $\begin{array}{c} (N,N) = (10,10)\\(N,N) = (12,12) \end{array}$ & \\
\hline
&Euclidean GRU & 60 & 11462 &  $(N,N) = (8,8)$ & \\
& & 70 & 15472 &  $\begin{array}{c} (N,N) = (10,10)\\(N,N) = (12,12) \end{array}$ & \\
\hline
&Poincar\'e RNN & 60 &  3902&  $(N,N) = (8,8)$ & \\
& & 70 & 5252  &  $\begin{array}{c} (N,N) = (10,10)\\(N,N) = (12,12) \end{array}$ & \\
\hline
&Poincar\'e GRU & 60 & 11462 &  $(N,N) = (8,8)$ & \\
& & 70 & 15472 &  $\begin{array}{c} (N,N) = (10,10)\\(N,N) = (12,12) \end{array}$ & \\
\hline

&Lorentz RNN & 60 & 3902 &  $(N,N) = (8,8)$ & \\
& & 70 & 5252  &  $\begin{array}{c} (N,N) = (10,10)\\(N,N) = (12,12) \end{array}$ & \\
\hline
&Lorentz GRU & 60 & 11462 &  $(N,N) = (8,8)$ & \\
& & 70 &  15472&  $\begin{array}{c} (N,N) = (10,10)\\(N,N) = (12,12) \end{array}$ & \\
\hline\hline
\end{tabular}
\caption{Six types of 1D Euclidean and hyperbolic Poincar\'e/Lorentz RNN/GRU NQS ansatzes used in the 2D TFIM VMC experiments. For $(N,N) = (8,8)$, all ansatzes have the hidden RNN/GRU dimension of 60, while for $(N,N) = (10,10)$, (12,12), the hidden dimension is 70. With the same hidden dimension, RNN variants have almost three times fewer parameters as GRU variants. } \label{1dnqs_detail}
\end{table}
\FloatBarrier
\begin{table}[!ht]
\centering
\begin{tabular}{c lccc c}
\hline\hline
 & Ansatz & $(N,N) =(8,8)$ &  $(N,N) =(10,10)$  & $(N,N) =(12,12)$ & \\
\hline\hline
& Euclidean RNN &  -201.4058 & -309.1322  & -447.3467 & \\
&& 0.0314 & 0.0780 &  0.0852 & \\
& Poincare RNN & \textbf{-201.7627} & -314.8574 &-445.6974 & \\
&& 0.0269 &  0.0517 &  0.0944 &\\
& Lorentz RNN & -201.5256 & \textbf{-315.1620} & -445.4792 &\\
&&  0.0315 & 0.0412 &  0.0938 &\\
& Euclidean GRU & -200.9939 &-314.1182 &-452.3138&\\
&& 0.0441 & 0.0611 & 0.0763&\\
& Poincare GRU & \textbf{-201.7624}  & \textbf{-315.3490} & \textbf{-454.2120}&\\
&& 0.0296 & 0.0449 & 0.0636&\\
& Lorentz GRU &  -200.8296 & -314.6966 & \textbf{-453.5092}&\\
&& 0.0445  & 0.0537 &  0.0691 &\\
\hline
& DMRG &  -202.5077  &-316.9770 & -457.0416 &\\
\hline
\end{tabular}
\caption{This table lists the \textit{best} results for each NQS ansatz (chosen from different runs corresponding to different random NQS weight initialization seeds) of 2D TFIM VMC experiments using 1D Euclidean and hyperbolic Poincar\'e/Lorentz RNN/GRU NQS ansatzes for different square lattice sizes $(N,N)$ at $B_x = 3.0$. The number of samples used for inference is $10^4$. For each NQS ansatz, we first list the mean energy in the first line, followed by the standard error (corresponding to VMC sampling) directly below in the second line. For each $(N,N)$ experiment setting, the two best performing NQS ansatzes are noted in bold. Note that in all three cases, the top performing ansatzes are always hyperbolic NQS. } \label{1dnqs_vmc_results}
\end{table}
\FloatBarrier

\begin{table}[!ht]
\centering
\begin{tabular}{c lccc c}
\hline\hline
 & Ansatz & $(N,N) =(8,8)$ &  $(N,N) =(10,10)$  & $(N,N) =(12,12)$ & \\
\hline\hline
& Euclidean RNN &  -197.8471  &-308.6596 & -444.7609& \\
&& 0.0528 & 0.0801 & 0.0889 & \\
& Poincare RNN & \textbf{-201.3019}  &-311.4692 & -444.7816 & \\
&& 0.0328 & 0.0705 & 0.0996 &\\
& Lorentz RNN &-200.5496  &-312.2401 &-445.2729 &\\
&& 0.0409 & 0.0637 & 0.0929 &\\
& Euclidean GRU & -200.9467 & -313.9553 & -452.0062&\\
&& 0.0441 & 0.0617  & 0.0772&\\
& Poincare GRU &\textbf{-201.6585} & \textbf{-314.9809} & \textbf{-453.8204}&\\
&& 0.0312 & 0.0506 &  0.0668&\\
& Lorentz GRU &-200.7991 & \textbf{-314.1184} & \textbf{-452.9004}&\\
&& 0.0457 & 0.0591 & 0.0735 &\\
\hline
& DMRG &  -202.5077  &-316.9770 & -457.0416 &\\
\hline
\end{tabular}
\caption{This table lists the \textit{averaged} results for each NQS ansatz across different runs (corresponding to different NQS seeds) of 2D TFIM VMC experiments using 1D Euclidean and hyperbolic Poincar\'e/Lorentz RNN/GRU NQS ansatzes for different square lattice sizes $(N,N)$ at $B_x = 3.0$. The number of samples used for inference is $10^4$. For each NQS ansatz, we first list the mean energy in the first line, followed by the standard error (corresponding to the VMC sampling) directly below in the second line. For each $(N,N)$ experiment setting, the two best performing NQS ansatzes are noted in bold. Note that in all three cases, the top performing ansatzes are always hyperbolic NQS.} \label{1dnqs_vmc_results_ave}
\end{table}
\FloatBarrier

From Table \ref{1dnqs_vmc_results} and Fig.\ref{1dnqs_vmc_ranking}, Table \ref{1dnqs_vmc_results_ave} and Fig.\ref{1dnqs_vmc_ranking_averaged}, as well as Figs.\ref{ss-corr-N8-1dnqs}, \ref{ss-corr-N10-1dnqs}, \ref{ss-corr-N12-1dnqs} the following observations are noted.
\begin{itemize}
  \item $(N,N)=(8,8)$:
  \bei
  \item In terms of the best results chosen from different VMC runs (see the first subfigure of Fig.\ref{1dnqs_vmc_ranking}), the best three performing NQS ansatzes  are Poincar\'e RNN (at $-201.7627$), Poincar\'e GRU  (at $-201.7624$) and Lorentz RNN ($-201.5256$). Interestingly, Euclidean RNN could reach a best value of -201.4058, outperforming the best reachable by Euclidean GRU ($-200.9939$) and Lorentz GRU ($-200.8296)$. The underperformance of Lorentz GRU in this case can be attributed to a sub-optimal selection of hyperparameters rather than the inherent ability of Lorentz GRU as an NQS ansatz.
  In terms of the performances averaged over different VMC runs (see top subfigure of Fig.\ref{1dnqs_vmc_ranking_averaged}), Poincar\'e GRU is the best NQS, followed by Poincar\'e RNN, each with the energy average in the range $(-201.70, -201.30)$. Euclidean RNN, however, is the worst ansatz, with an energy average of $-197.8471$, underperforming both Lorentz GRU and Euclidean GRU.
  \item
  In terms of the spin-spin correlation decay $\langle \s^z_0\s^z_r\rangle$, Fig.\ref{ss-corr-N8-1dnqs}, all curves show a peak of correlation at $r=8$, signaling their abilities to capture the pattern of 2D-1D mapping Eq.(\ref{2d-1d-map}) when the 2D lattice size is $N=8$.
   Among the RNN-based variants,  Euclidean RNN has the flattest curve, followed by Lorentz RNN and Poincare RNN. The GRU-based variants have almost identical curves which all show a clear periodic pattern that is much more pronounced than the RNN-based variants, demonstrating their superior ability to capture the hierarchical interaction patterns $\langle i, i+1\rangle$ - $\langle i, i+8\rangle$. Note that while the 3D CFT reference curve shows a relatively flat decay at large distance, the more accurate picture is one in which the correlation drops sharply near the edges due to the open boundary conditions. This is what is observed for the curves produced by the GRU variants, which accurately capture this sharp decay at large distances.
  \eni
  \item $(N,N)=(10,10)$:
    \bei
  \item In terms of the best value reachable by each NQS (see the second subfigure of Fig.\ref{1dnqs_vmc_ranking}), Poincar\'e GRU ($-315.3490$) is the best NQS, followed by Lorentz RNN ($-315.1620$) and Poincar\'e RNN ($-314.8574$) and Lorentz GRU at the fourth place.  In terms of the averaged results over different VMC runs (see the second subfigure of Fig.\ref{1dnqs_vmc_ranking_averaged}), the best performing ansatzes are Poincar\'e GRU, Lorentz GRU and Euclidean GRU.
\item
In terms of the spin-spin correlation $\langle \s^z_0\s^z_r\rangle$, Fig.\ref{ss-corr-N10-1dnqs}, all curves, except Euclidean RNN's, show a peak at $r=10$. However, the curves corresponding to the GRU-based variants (which are again almost identical among the three variants of Euclidean, Poincar\'e and Lorentz GRU) are much more periodic than the RNN-based ones which are almost flat by comparison. This again signals the better ability of the GRU variants at capturing the 2D-1D mapping Eq.(\ref{2d-1d-map}) and interaction hierachy $\langle i,i+1\rangle$ - $\langle i,i+10\rangle$.
\eni
  \item $(N,N)=(12,12)$:
    \bei
  \item In terms of best energy value chosen from different runs, Poincar\'e GRU ($-454.2120$) is again the best performing ansatz, followed by Lorentz GRU ($-453.5092$), and Euclidean GRU ($-452.3138$), far exceeding the RNN variants where Euclidean RNN (at $-447.3467$) is the best performing variant. In terms of the averaged performances across different runs (Table \ref{1dnqs_vmc_results_ave}), the top three spots remain exactly the same, but among the bottom three spots, Lorentz RNN ranks first, followed by Poincar\'e RNN and Euclidean RNN. It is interesting to note that among the RNN variants, Euclidean RNN actually outperforms Lorentz and Poincar\'e RNN in terms of best $E$ reachable, but underperforms both in terms of averaged $E$ reachable. The underperformances of Lorentz and Poincar\'e RNN do not signal their lack of inherent expressivity as NQS ansatzes but rather are due to a lack of comprehensive hyperparameter tuning to select the optimal configurations.
  \item
  In terms of the spin-spin correlation decay in Fig.\ref{ss-corr-N12-1dnqs}, the three NQS in the RNN variants fail completely to capture the $12^{th}$ neighbor interaction, as their curves are completely flat while the GRU variants still successfully capture this 2D-1D mapping in this case. Among the GRU variants, hyperbolic Poincar\'e and Lorentz GRU outperform Euclidean GRU, as evidenced by their corresponding curves - the hyperbolic ones clearly show the periodic structure with the $r=12$ period, even at large distances while the Euclidean one flattens out quickly with distance. In particular, at intermediate distances ($20 < r < 70$), Poincaré and Lorentz GRUs trace both the peaks and the valley depths of the translated 3D CFT baseline with high fidelity, whereas Euclidean GRU fails to decay deeply enough in the valleys due to an unphysical long-range correlation tail. At extreme distances ($r > 80$), the Euclidean curve appears to match the continuum CFT line better, but this is a known artifact of under-optimization. The hyperbolic GRUs correctly capture the finite-size boundary effect by plunging at the tail, while the Euclidean network over-estimates long-range alignment.
  \eni
  \item Overall, across three different metrics (absolute best value reachable, averaged performance, spin-spin correlation decay length), at three different lattice sizes $(N,N)$=(8,8),(10,10), (12,12), hyperbolic NQS ansatzes consistently outperform their Euclidean versions.  In particular,  Poincar\'e GRU emerges as the best performer for all three 2DTFIM settings, outperforming all other NQS ansatzes, including Euclidean GRU and Lorentz GRU. This observation agrees with and extends our previous result from \cite{hld-hypnqs-25} where we showed that Poincar\'e GRU consistently outperforms Euclidean GRU NQS with different lattice sizes from ($N,N$) = (5,5) to (9,9). In this work, as the lattice size increases from $(8,8)$ to $(10,10)$ to $(12,12)$, the GRU variants increasingly demonstrate their superior abilities over the RNN variants to capture the ever-larger hierarchical $N^{th}$ neighbor interaction where $N$ goes from 8, 10, to 12 as seen from the series of spin-spin correlation curves in Figs.\ref{ss-corr-N8-1dnqs}, \ref{ss-corr-N10-1dnqs}, \ref{ss-corr-N12-1dnqs}. Among the GRU variants, as $N$ increases to 12, hyperbolic Lorentz and Poincar\'e GRU show a clear dominance over Euclidean GRU.
  \\\\
  While the probable main reason for the clear outperformance achieved by 1D hyperbolic NQS over Euclidean ones in this unrolled 2DTFIM setting is the presence of the structural hierarchy $\langle i,i+1\rangle$-$\langle i,i+N\rangle$ of neighbor interactions, one must also not discount the fact that these VMC experiments are performed at the critical magnetic field strength $B_x=3.0$ where the system displays CFT physics. As discussed in the previous section, within the context of our physically motivated heuristic hypothesis, the connection to the AdS/CFT correspondence could also be a determining factor in the observed outperformance by hyperbolic 2D NQS over Euclidean 2D NQS in the lack of structural hierarchy. In this case, we have both structural hierarchy and CFT physics which can both contribute to the advantage of hyperbolic 1D NQS models. Furthermore, we hypothesize that if the VMC experiments in the unrolled 2DTFIM setting were to be done at $B_x=2.0$, 1D hyperbolic NQS ansatzes would still outperform their 1D Euclidean NQS counterparts despite the lack of a CFT physics description because the long range, ordered nature of the 2D ground state at $B_x=2.0$ demands that this ordered correlation be preserved in the hierarchical structure $\langle i,i+1\rangle$-$\langle i,i+N\rangle$ in the 1D setting. For $B_x=4.0$, it is less certain that 1D hyperbolic NQS models would outperform their 1D Euclidean NQS versions in the unrolled 2DTFIM because the disordered nature of the 2D ground state at $B_x=4.0$ weakens the $\langle i,i+1\rangle$-$\langle i,i+N\rangle$ structure.

\end{itemize}
\FloatBarrier
\FloatBarrier
\FloatBarrier
\FloatBarrier

\FloatBarrier

\section{Concluding remarks} \label{concl}
In this work, we introduced the Lorentz 2DRNN NQS - the first two-dimensional non-Euclidean NQS construction. We benchmarked the performances of Lorentz 2DRNN against its Euclidean counterpart, the two-dimensional Euclidean 2DRNN NQS - a custom construction introduced in \cite{rnn_20} in a variety of different 2DTFIM VMC experiment settings involving different lattice sizes ($N,N$) at different magnetic field strengths $B_x$. In addition to the newly constructed two-dimensional hyperbolic Lorentz 2DRNN, we also expanded the results of \cite{hld-hypnqs-25} concerning the use of one-dimensional hyperbolic NQS in the 2DTFIM setting where the two dimensional system is translated/unrolled into one dimension. In this work, the 1D NQS ansatzes considered include not only the original Poincar\'e GRU first introduced in \cite{hld-hypnqs-25}, but also new 1D hyperbolic NQS such as Lorentz RNN, Lorentz GRU and Poincar\'e GRU introduced in \cite{hld-hypnqs-26}.
\\\\
Our main findings are summarized below.
\begin{itemize}
  \item  \textit{Regarding two-dimensional NQS ansatzes}: Within the scope of the native 2DTFIM experiments performed, Lorentz 2DRNN NQS robustly outperforms Euclidean 2DRNN in all three system sizes  $8\times 8$, $10\times 10$, $12\times 12$ at $B_x=3.0$ (at the phase transition point) both in terms of best and averaged reachable mean ground state energy (see Table \ref{2dtfim_varied_N8_res}, Table \ref{2dtfim_varied_N10_res}, Table \ref{2dtfim_varied_N12_res}). In terms of the spin-spin correlation $\langle\s^z_0\s^z_r\rangle$ in the case of $12\times 12$ 2DTFIM (see Fig.\ref{fig-spin-corr-N12-ave}), the power-law correlation decay is best reproduced by a Lorentz 2DRNN NQS with $L_{max} = 5.0$ among the three NQS ansatzes considered: Euclidean 2DRNN, Lorentz 2DRNN with $L_{max}=2.0$ and Lorentz 2DRNN with $L_{max}=5.0$.  Away from the phase transition point, when $B_x=2.0$, Euclidean 2DRNN NQS robustly outperforms hyperbolic Lorentz 2DRNN (see Table \ref{2dtfim_varied_B2}), while for $B_x=4.0$, the performances of the two are comparable (see Table \ref{2dtfim_varied_B4}).
  \\\\
   As described in detail in Section \ref{sub:varied_N_fixed_B}, a physically motivated heuristic hypothesis for the observed outperformance of Lorentz 2DRNN compared to Euclidean 2DRNN at the phase transition point involves the fact that the physics of the 2DTFIM system at criticality is described by a conformal field theory (CFT) that is known to be dual to an AdS space whose spatial geometry is none other than hyperbolic space. Thus, the hyperbolic space underlying the construction of the Lorentz 2DRNN naturally endows it with a favorable geometric inductive bias and a representational capacity that are well-matched to the physics of the critical system. This is reminiscient of the way another non-NQS variational ansatz, MERA (Multi-scale Entanglement Renormalization Ansatz), is known to  efficiently represent quantum states at criticality \cite{vidal-06}, \cite{swingle-09}, \cite{swingle-12}: MERA, with its exponential tree-like structure of isometries and disentanglers is fundamentally a discrete version of hyperbolic space whose geometry matches the CFT physics of the critical quantum states. While both our hyperbolic NQS and MERA share a hyperbolic geometry advantage when it comes to representing critical quantum states, their main difference, apart from the obvious fact that one is neural-network-based and one is tensor-based, lies in the continuous versus discrete nature of their hyperbolic constructions. As MERA builds a discrete, layer-by-layer hyperbolic tree to represent a state, our hyperbolic NQS utilizes a continuous hyperbolic manifold to achieve a structurally parallel representation of critical spatial scaling. 
  \item \textit{Regarding one-dimensional NQS ansatzes}:  1D hyperbolic NQS ansatzes robustly outperform their Euclidean counterparts in all three lattice sizes considered $(N,N)$ = (8,8), (10,10), (12,12) at $B_x=3.0$ (see Table \ref{1dnqs_vmc_results} and Table \ref{1dnqs_vmc_results_ave}, as well as Fig.\ref{1dnqs_vmc_ranking} and Fig.\ref{1dnqs_vmc_ranking_averaged}). In terms of the mean energy value, Poincar\'e GRU emerges as the best overall NQS ansatz in all three cases under study, with the second best ansatz varying in each case: Lorentz GRU for $N=12$, Lorentz RNN for $N=10$ and Poincar\'e RNN for $N=8$.
\\\\
  The use of 1D NQS necessitates the conversion of the $N\times N$ 2DTFIM into a one-dimensional setting where the vertical nearest neighbors in the 2D lattice become the $N^{th}$ neighbor in the 1D spin chain, which means that a different lattice size $N$ leads to a different type of hierarchical structure such that the larger $N$ is, the harder it becomes for the 1D NQS ansatz to capture this structure. This is clearly seen in the spin-spin correlation length $\langle\s^z_0\s^z_r\rangle$ plots, Fig.\ref{ss-corr-N8-1dnqs}, Fig.\ref{ss-corr-N10-1dnqs} and Fig.\ref{ss-corr-N12-1dnqs}, where a successful 1D NQS is one that displays the periodic structure in $\langle \s^z_0\s^z_r\rangle$ with the  period corresponding to the lattice size $N$.
  When $(N,N)$ = (8,8), all ansatzes - including 1D Euclidean RNN - display the periodic pattern in their curves with the period of $r=8$. When $(N,N)$=(10,10), all ansatzes - excluding Euclidean RNN - display the periodic pattern with the period of $r=10$.  The GRU variants display a much more pronounced periodic structure in their correlation curves than the RNN variants, whose curves are far flatter by comparison, for both these system sizes. However, as the system size increases to $(N,N)$=(12,12), only the GRU variants display the periodic pattern with the period of $r=12$, while the RNN variant curves are completely flattened out. Even among the three GRU variants, the Euclidean GRU curve is noticeably flatter compared to the Poincar\'e's and Lorentz GRU's.
  \\\\
  Apart from the obvious structural hierarchy $\langle i, i+1\rangle$-$\langle i,i+N\rangle$ in the neighbor interactions that can account for the outperformance of 1D hyperbolic NQS compared to their Euclidean counterparts as pointed out in \cite{hld-hypnqs-25}, in this work, we also note that the presence of the CFT physics at the critical point where the VMC experiments are carried out also plays an important part. In particular, within the context of our AdS/CFT heuristic hypothesis as outlined above, it might be a combination of both these factors (structural hierarchy and CFT physics) that contributes to the observed robust outperformance of 1D hyperbolic NQS ansatzes at the critical point in the unrolled 2DTFIM setting. While we did not perform the VMC experiments for the unrolled 2D setting at other magnetic field strengths besides $B_x=3.0$, we hypothesize that at $B_x=2.0$ when the ground state is an ordered ferromagnet, in the absence of CFT physics, 1D hyperbolic NQS models would continue to outperform their Euclidean versions thanks to the fact that the ordered nature of the state can still be captured effectively through structural hierarchy. For $B_x=4$, when the system is in a disordered state, it is uncertain whether 1D hyperbolic NQS would have an advantage over 1D Euclidean NQS since structural hierarchy does not play a role in conveying the disordered nature of the system.
\end{itemize}
While further numerical studies involving larger system sizes and formal mathematical proofs remain necessary to fully validate our proposed geometric hypothesis, our benchmark results demonstrate that hyperbolic geometric inductive biases provide a robust advantage for quantum state representation within the scope of the conducted VMC experiments. Ultimately, this work serves as a proof-of-concept for the enhanced expressivity and superior capture of long-range CFT correlation decay by hyperbolic NQS models near the critical point. Given the results of this work, several interesting future directions emerge. 
\bei
\item As already noted in \cite{hld-hypnqs-25}, \cite{hld-hypnqs-26}, the direction of immediate interest might be the new construction of the already-proposed 1D/2D hyperbolic NQS\footnote{In our series of works on hyperbolic NQS so far, we have always used a fixed curvature constant of $c=1$ for the Poincar\'e networks and $k=1$ for the Lorentz networks.} in this work and \cite{hld-hypnqs-25}, \cite{hld-hypnqs-26} based on a learnable or trainable curvature parameter \cite{graph-classification} for both the Poincar\'e disk and the Lorentz hyperboloid, which might remove the need to explicitly enforce spatial constraint hyperparameters $L_{max}$ and $R_{max}$. Furthermore, it would be crucial if more efficient and systematic hyperparameter tuning methods for hyperbolic NQS could be implemented so that optimal hyperparameters can be selected rather than relying on a trial-and-error basis as was done in our work, due to a lack of computational resources.

\item Another immediate direction might involve the extension of the Lorentz 2DRNN construction in this work to the more complex 2DGRU version or even more broadly, exploring other constructions of both 1D and 2D hyperbolic NQS ansatzes based on completely different types of architectures, such as purely feedforward neural network,  CNN (Convolutional Neural Network), hybrid CNN-RNN\footnote{An example of this hybrid CNN-RNN architecture was proposed in the work  \cite{hld-cicy-rnn} in the context of predicting the topological Hodge numbers of Calabi-Yau four-folds for string theory compactification applications.} or transformers, and applying these new constructions to the paradigmatic settings such as the TFIM or Heisenberg models. 
\\\\
Along this direction, our recent work \cite{hrbm-2609} moves beyond hyperbolic recurrent architectures to introduce the first non-autoregressive non-Euclidean NQS in the form of the hyperbolic restricted Boltzmann machine (HRBM) and evaluates its performance against the Euclidean RBM in the setting of the quantum Sherrington-Kirkpatrick (QSK) model whose ground state exhibits volume-law entanglement. The results of \cite{hrbm-2609} show that HRBM robustly outperforms RBM in terms of the lower relative energy error, lower variance, as well as lower R\'enyi-2 and von Neumann reconstruction errors across all tested QSK system sizes spanning a 512-fold increase in the Hilbert space dimension. More importantly, thanks to the exponential volume of the hyperbolic latent space underlying HRBM's construction, which naturally mirrors the hierarchical entanglement structure of the volume-law ground state, HRBM faithfully reproduces the full entanglement spectra of all QSK disorder realizations under study across 15 orders of magnitude from the top eigenvalues down to sub-dominant modes, which RBM consistently over-estimates. This work marks a key step in demonstrating that the superior expressivity and representational advantages of hyperbolic NQS models extend beyond recurrent architectures to generalize across broader neural quantum state paradigms.
\item
Going farther afield beyond the context of spin models, it might also be interesting and instructive to explore the performances of hyperbolic NQS in different quantum systems\footnote{Incidentally, the work \cite{vscore-23} provides a comprehensive list of the different quantum many-body systems and the different variational methods used to study them.}, especially those with known AdS holographic duals - one example of which is the $SU(N)$ matrix models that have recently been explored in the context of variational quantum circuits \cite{hld-qc}, \cite{qc-mm}. 
\eni
\newpage
\section{Appendix} \label{sec:appendix}
\subsection{Hierarchical structures of neighbor interactions in spin systems}\label{graph-forms}
\begin{figure}[!ht]
\centering
\includegraphics[width = .5\textwidth]{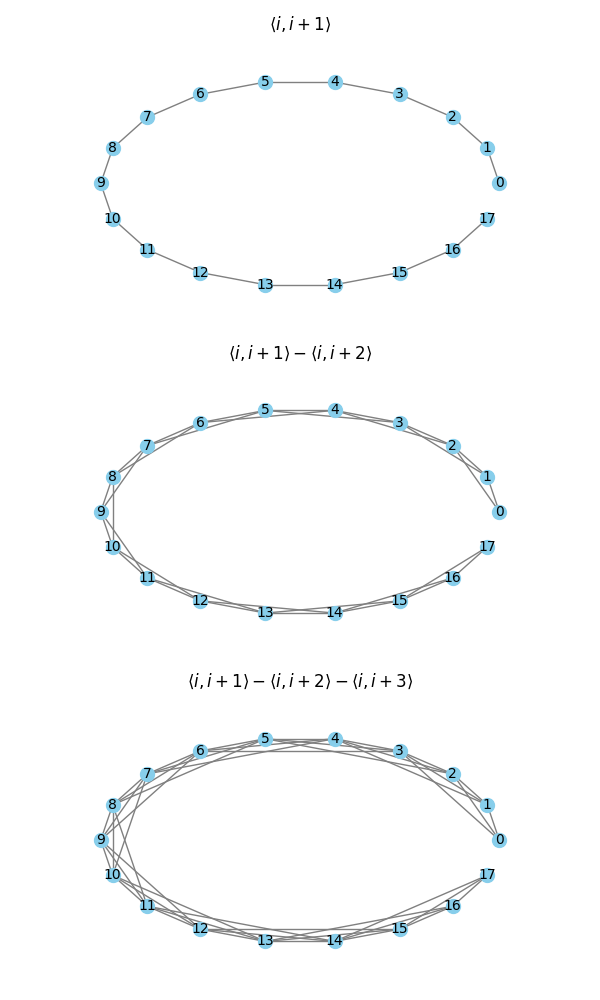}
\caption{An illustration of the  interaction hierarchies of different degrees of nearest neighbor interactions in the form of graphs in spin models for 1D Heisenberg $J_1J_2$ and $J_1J_2J_3$ spin chain models. For illustration purpose, the number of sites is chosen to be $N=18$. }\label{graph-like-h}
\end{figure}

\begin{figure}[!ht]
\centering
\includegraphics[width = .5\textwidth]{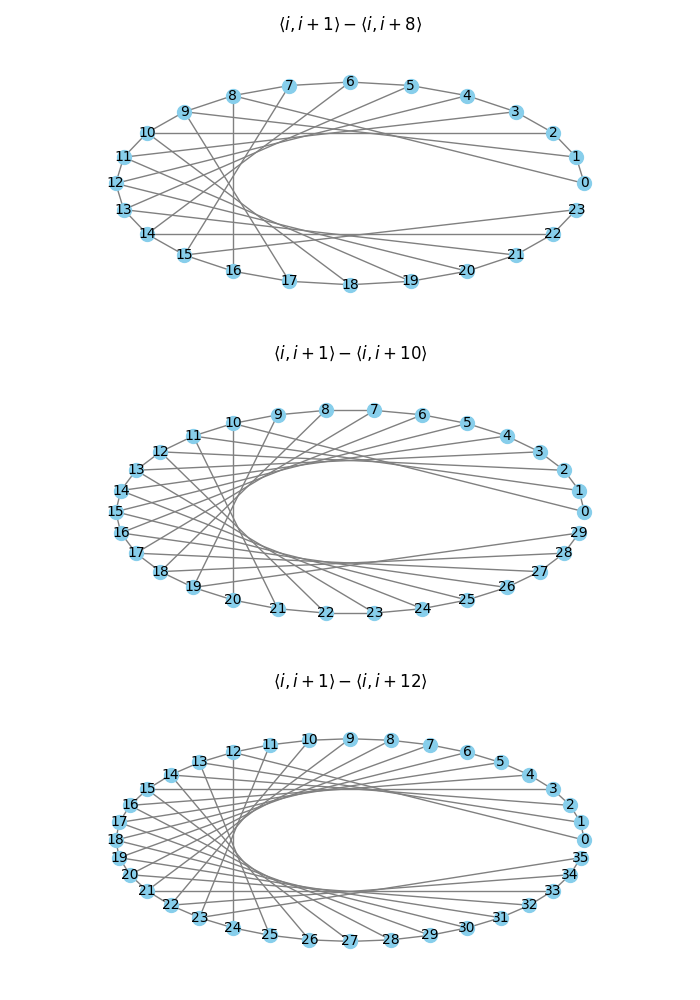}
\caption{An illustration of the interaction hierarchies of different degrees of nearest neighbor interactions in the form of graphs for $8\times 8$, $10\times 10$ and $12\times 12$ 2DTFIM  mapped to 1DTFIM spin chain model where the vertical nearest neighbor in the  $N\times N$ lattice becomes the $N^{th}$ neighbor in the 1D spin chain. For illustration purpose, the number of sites is chosen to be 24, 30 and 36.  }\label{graph-like-t}
\end{figure}

In this section, we illustrate the hierarchical neighbor interactions of different spin systems including TFIM and Heisenberg $J_1J_2$ and $J_1J_2J_3$ systems in the form of graphs in Fig.\ref{graph-like-h} and Fig.\ref{graph-like-t}.
\\\\
 Each type of neighbor interaction, be it first ($\langle i, i+1\rangle$), second ($\langle i, i+2\rangle$), third ($\langle i, i+3\rangle$) or $N^{th}$ ($\langle i, i+N\rangle$), is represented by an edge connecting vertex $i$ and vertex $j$ (where $j=1,2,3,N$) in a graph whose vertices are the spin sites. In this manner, the neighbor interaction structures corresponding to the Heisenberg $J_1J_2$ and $J_1J_2J_3$ spin systems are shown in Fig.\ref{graph-like-h}, while the neighbor interaction structures corresponding to the 2DTFIM mapped to the 1DTFIM using Eq.\ref{2d-1d-map} are shown in Fig.\ref{graph-like-t}. In this case,  the 2D lattice sizes $(N,N) = (8,8)$, (10,10), (12,12) correspond to the $8^{th}$ - $\langle i, i+8\rangle$, $10^{th}$ - $\langle i, i+10\rangle$ and $12^{th}$ - $\langle i, i+12\rangle$ neighbor interactions in the 1D spin chain.
 \\\\
 Collectively, these graphs can be very loosely categorized as Watts-Strogatz \cite{watts-strogatz} type, or `small-world' networks consisting of vertices that are connected to a fixed number of neighboring vertices to form a network. This model is capable of interpolating between regular and random networks by using a certain nonzero `rewiring' probability $p$ that randomly rewires the connections between vertices. In our context, the neighbor interaction structures of spin models correspond to Watts-Strogatz graphs with zero rewiring probability.
 Incidentally, in the graph-embedding and graph classification context, hyperbolic neural networks have been shown to outperform their Euclidean counterparts in these tasks thanks to their more efficient representational capacity \cite{graph-classification}.
\FloatBarrier

\FloatBarrier
\subsection{Poincar\'e disk model of hyperbolic space} \label{ssub:poincar_disk}
The Poincar\'e ball model $(\mathbb{D}^N, g^{\mathbb D})$ of hyperbolic space is defined by the manifold $\mathbb{D}^N$
\beq
\mathbb{D}^N_c = \left\{ x\in \mathbb{R}^N: c||x||<1\right\} \label{eq-gyro}
\eeq
where the parameter $c$ is the Poincar\'e ball's radius. When $c=0$, $\mathbb{D}^N_c = \mathbb{R}^N$, while when $c>0$, $\mathbb{D}^N_c$ is the open ball of radius $1/\sqrt{c}$. In all the computations used in this work, we set $c=1$. 
The space $\mathbb{D}^N_c$ in Eq.(\ref{eq-gyro}) is equipped with the metric
\beq
g^{\mb D}_x = \lambda^2_x g^E, \hspace{5mm} \lambda_x = \frac{2}{1-||x||^2} \label{eq:poincare-metric}
\eeq
where $g^E = \mathbf{1}_N$ is the identity matrix representing the Euclidean metric. 
The parallel transport $P^c_{\mf{0}\ra x}(v)$ of a vector $v\in T_\mf{0}\mb{D}^N_c$ to another tangent space $T_x\mb{D}^N_c$ are defined as
\beq
P^c_{\mf{0}\ra x}(v) = \log_x^c\lf[x\oplus_c \exp_\mf{0}^c(v)\rr] \label{eq-pt}
\eeq
where the operations $\log$ and $\exp$ are the exponential and logarithmic maps. For any point $x\in \mb{D}^N_c$, any vector $v\neq \mf{0}$ and any point $y\neq x$, the exponential and logarithmic maps $\exp^c_x: T_x\mb{D}^M_c \ra \mb{D}^M_c $ and $\log^c_x: \mb{D}^N_c \ra T_x\mb{D}^N_c$ between the hyperbolic space and its Euclidean tangent space are defined as
\beq
\exp_x^c(v) &=& x \oplus_c\lf[ \tanh\lf(\sqrt{c}\frac{\lambda_x^c||v||}{2}\rr)\frac{v}{\sqrt{c}||v||}\rr] \label{eq-exp}
\\
\log_x^c(y) &=& \frac{2}{\sqrt{c}\l_x^c} \tanh^{-1}\lf(\sqrt{c}||-x\oplus_c y||\rr) \frac{-x\oplus_cy}{||-x\oplus_c y||} \label{eq-log}
\eeq
where $\l_x^c = 2/(1-c||x||^2)$. When $x = \mf{0}$, the above maps take  more compact forms
\beq
\exp_\mf{0}^c(v) &=& \tanh\lf(\sqrt{c}||v||\rr) \frac{v}{\sqrt{c}||v||}\,\,\qquad 
\lf(T_{\mf{0}_M}\mb{D}^M_c \ra \mb{D}^M_c \rr) \label{eq-exp0}
\\
\log_\mf{0}^c(y) &=& \tanh^{-1}\lf(\sqrt{c}||y||\rr) \frac{y}{\sqrt{c}||y||}\qquad \lf(\mb{D}^N_c \ra T_{\mf{0}_N}\mb{D}^N_c\rr)\,. \label{eq-log0}
\eeq
The mathematical operations $\oplus_c, \otimes_c, \odot_c$ appearing in the equations above are the Poincar\'e hyperbolic analogs of the Euclidean addition, matrix multiplication and pointwise multiplication.
\bei
\item For $x,y \in \mb{D}^N_c$, the Mobius addition $\oplus_c$ is
\beq
x\oplus_c y \equiv \frac{(1+2c\langle x, y\rangle + c||y||^2)x + (1-c||x||^2)y}{1 + 2c \langle x, y\rangle + c^2 ||x||^2 ||y||^2}\,. \label{eq-oplus}
\eeq
In terms of parallel transport, the Mobius addition for $x\in \mb{D}^N_c$ with $b\in \mb{D}^N_c$ can be written as
\beq
x\oplus_c b = \exp_x^c\lf[P_{\mf{0} \ra x}^c\lf(\log_\mf{0}^c(b)\frac{}{}\rr)\rr]\label{eq-m-pt}\,.
\eeq
\item For $x\in \mathbb{D}^N_c\backslash \{\mathbf{0}\}$ and $W: \mathbb R^M \ra \mathbb R^N$, the Mobius matrix multiplication is defined by
\beq
 W \otimes_c x = \frac{1}{\sqrt{c}} \tanh\lf(\frac{||Wx||}{||x||} \tanh^{-1}(\sqrt{c}||x||)\rr)\frac{Wx}{||Wx||}
\eeq
\item For $x\in \mathbb{D}^N_c\backslash \{\mathbf{0}\}$ and $r\in \mathbb R$, the Mobius pointwise multiplication $\odot_c$, is defined as
\beq
r\odot_c x = \frac{1}{\sqrt{c}} \tanh\lf(\frac{||rx||}{||x||} \tanh^{-1}(\sqrt{c}||x||)\rr)\frac{rx}{||rx||}
\eeq
\item The nonlinear activation $f^{\otimes_c}$(x) where $x\in \mb D^N_c$ is
\beq
f^{\otimes_c} = \exp^c_{\mf 0}\lf(f(\log^c_{\mf 0}(x))\rr)
\eeq
\eni

\subsection{Lorentz hyperboloid model of hyperbolic space}\label{ssub:lorentz_model}
The $n$-dimensional Lorentz hyperboloid $\mathbb H^n$ with constant negative curvature $-k$ where $k=1$ is given as
\beq
\mathbb H^n \equiv \left\{\mathbf x\in \mathbb R^{n+1}: \langle \mathbf x,\mathbf x\rangle_{\mathcal L} = -1, \,x_0>0\rr\}
\eeq
where $\langle \mathbf x, \mathbf y \rangle_{\mathcal L}$, the Lorentzian scalar product for $(n+1)$-dimensional vectors $\mathbf{x, y} \in \mathbb R^{n+1}$, is defined as
\beq
\langle \mathbf x, \mathbf y \rangle_{\mathcal L} \equiv -x_0y_0 + \sum_{i=1}^n x_iy_i\,.
\eeq
Unlike the Poincar\'e disk where the origin $\mathbf 0_P = (0,0,\ldots, 0)$, in the Lorentz model of hyperbolic space, the origin is the point $\mathbf 0_{\mathcal L} = (1, 0, 0,\ldots, 0)$.
The distance function between two points $\mathbf x, \mathbf y \in \mathbb H^n$ is 
\beq
d_{\mathbb H}(\mathbf x, \mathbf y) = \text{arcosh}\left(-\langle  \mathbf x, \mathbf y\rangle_{\mathcal L}\rr)\,,
\eeq
while the Lorentzian norm of a vector $\mathbf v$ is 
\beq
||\mathbf v||_{\mathcal L} = \sqrt{\langle  \mathbf v, \mathbf v\rangle_{\mathcal L}}\,.
\eeq
The tangent space at $\mathbf x$ is the $n$-dimensional Euclidean vector space approximating $\mathbb H^n$ around $\mathbf x$:
\beq
\mathcal T_x\mathbb H^n \equiv \lf\{ \mathbf x\in \mathbb R^{n+1}: \langle \mathbf v, \mathbf x\rangle_{\mathcal L} = 0\rr\}
\eeq
Similar to the Poincar\'e model described in the previous section, mappings between the Lorentz hyperboloid and its tangent space are done using the exponential $\exp_{\mathbf x}(\mathbf v)$ and logarithmic $\log_{\mathbf{x}}(\mathbf y)$ maps.
\beq
\mathcal T_{\mathbf x} \mathbb H^n \rightarrow \mathbb H^n: &&\exp_{\mathbf x}(\mathbf v) = \cosh\lf(||\mathbf v||_{\mathcal L}\rr) \mathbf x + \sinh\lf(||\mathbf v||_{\mathcal L}\rr)\frac{\mathbf v}{||\mathbf v||_{\mathcal L}} \label{eq-l-expmap}\\
\mathbb H^n \rightarrow \mathcal T_{\mathbf x} \mathbb H^n: && \log_{\mathbf{x}}(\mathbf y) =d_{\mathbb H}(\mathbf x, \mathbf y) \frac{\mathbf y + \langle \mathbf x, \mathbf y\rangle_{\mathcal L}\,\mathbf x}{\left||\mathbf y + \langle \mathbf x, \mathbf y\rangle_{\mathcal L}\,\mathbf x\right||_{\mathcal L}} \label{eq-l-logmap}
\eeq
The parallel transport operation that maps a point $\mathbf{z} \in \mathcal T_{\mathbf x} \mathbb H^n$ to a point in $\mathcal T_{\mathbf y}\mathbb H^n$ is defined as
\beq
P_{\mathbf{x}\ra \mathbf{y}}(\mathbf{z}) = \mathbf z+\frac{\langle \mathbf y, \mathbf z\rangle_{\mathcal L}}{1-\langle(\mathbf x, \mathbf y)\rangle_{\mathcal L}}(\mathbf x+ \mathbf y)
\eeq
When $\mathbf x= \mathbf 0_{\mathcal L}$, 
\beq
P_{\mathbf 0_{\mathcal L} \ra \mathbf{y}}(\mathbf z) = \mathbf z+\frac{\langle \mathbf y, \mathbf z\rangle_{\mathcal L}}{1-\langle(\mathbf 0_{\mathcal L}, \mathbf y)\rangle_{\mathcal L}}(\mathbf 0_{\mathcal L}+ \mathbf y) \label{ptrans0}
\eeq
The various Lorentz mathematical operations appearing Eqs.(\ref{eq-lorentz-rnn}), (\ref{eq-lorentz-gru}) are defined in terms of the exponential/logarithm mappings at $\mathbf x= \mathbf 0_{\mathcal L}$ in (\ref{eq-l-expmap}), (\ref{eq-l-logmap}) and parallel transport operations given in (\ref{ptrans0})  as follows.
\bei
\item For $\mathbf x, \mathbf y \in \mathbb H^n$, the Lorentz addition $ \oplus_{\mathcal L}$ is defined as:
\beq
\text{Lorentz addition}: \qquad \mathbf x \oplus_{\mathcal L} \mathbf y = \exp_{\mathbf x}\lf[P_{\mathbf 0_{\mathcal L} \ra \mathbf{x}}(\log_{\mathbf 0_{\mathcal L}}(\mathbf y))\rr]
\eeq
\item The scalar multiplication $\odot_{\mathcal L}$ between a point $\mathbf x\in \mathbb H^n$ and $r\in \mathbb R$ is 
\beq
\text{Lorentz scalar multiplication}: \qquad r \odot_{\mathcal L} \mathbf x =\exp_{\mathbf 0_{\mathcal L}}\lf[ r \log_{\mathbf 0_{\mathcal L}}(\mathbf x)\rr]
\eeq
\item The matrix multiplication $\otimes_{\mathcal L}$ between a point $\mathbf x\in \mathbb H^n$ and a matrix  $M\in \mathbb R^n\times \mathbb R^n$ is 
\beq
M \otimes_{\mathcal L} \mathbf x =\exp_{\mathbf 0_{\mathcal L}}\lf[ M \log_{\mathbf 0_{\mathcal L}}(\mathbf x)\rr]
\eeq
\item The nonlinear activation $f^{\otimes_{\mathcal L}}$(x) where $x\in \mb H^n$ is
\beq
f^{\otimes_{\mathcal L}} = \exp_{\mathbf 0_{\mathcal L}}\lf(f(\log_{\mathbf 0_{\mathcal L}}(x))\rr)
\eeq
\eni
\FloatBarrier
\clearpage
\subsection{Training details and hyperparameters in all VMC experiments}
\label{ssub:hyperparams}
In this section, we describe the training details of all VMC experiments, which belong to the two different experiment settings listed below.
\ben
\item The first experiment setting involves  2D RNN NQS ansatzes (Euclidean and Lorentz) using the native two dimensional geometry of the 2DTFIM lattice. The experiments in this setting are labeled $\left([N,N], B_x\right)$ where $B_x=2.0, 3.0, 4.0$, (see Table \ref{tfim2d-hyperparams}),
\item The second experiment setting involves 1D RNN/GRN NQS ansatzes (Euclidean, Poincar\'e and Lorentz) using the unrolled geometry of the 2DTFIM lattice. The experiments in this setting are labeled by $([N,N])$ as $B_x$ is fixed to be 3.0 (see Table \ref{tfim2d-ur-hyperparams}).
\enn
The following details are common to all VMC experiments (in both settings listed above)\footnote{Many of these details are similar or identical to our recent works \cite{hld-hypnqs-25}, \cite{hld-hypnqs-26}}.
\bei
\item \textit{Number of VMC samples}: For training, we used 80 samples for all NQS ansatzes, while for inference with trained NQS, we use $10^4$ samples. The small number of samples used during training is justified by the fact that these are autoregressive NQS, which allows for exact and independent sample generations, unlike non-autoregressive NQS ansatzes utilizing Markov-chain Monte Carlo sampling, which typically requires thousands of samples. 

\item \textit{DMRG benchmark}: Although DMRG is used as a common benchmark for all NQS models studied in this work, DMRG is known to be not exact in two dimensions, unlike in one dimensional quantum systems where DMRG is considered to be exact. However, it is still a reliable enough benchmark in 2D when the bond dimension $\chi$ is chosen to be a large number, e.g. $\chi=512$ or $\chi=1024$, as done in \cite{rnn_20}. In fact, it is a common practice to use DMRG benchmarks in 2D systems (see, for example \cite{rnn_20}, \cite{rnn_22}, \cite{rnn_24}). For future studies, a good alternative to DMRG that can potentially serve as a more reliable benchmark in two dimensions is PEPS, which has been shown to outperform DMRG in computing ground states of 2D TFIM and Heisenberg models  \cite{peps-1}, \cite{peps-2}. 

\item \textit{Hyperbolic hyperparameters}: As described in detail in \cite{hld-hypnqs-26}, both Lorentz and Poincar\'e hyperbolic NQS models take much longer and are more complicated to train than Euclidean NQS, because their defining mathematical operations are much more complex and there are additional hyperparameters that have to be taken into account when dealing with hyperbolic networks. 
These additional hyperparameters are:
\bei
\item The hyperbolic learning rate (LR) that applies only to hyperbolic parameters in hyperbolic NQS models\footnote{Recall that hyperbolic neural networks have both Euclidean and hyperbolic parameters - see Section \ref{sec-phyprnn}}).  
\item The hyperbolic spatial constraint parameter, which is $R_{max}$ for Poincar\'e networks and $L_{max}$ for Lorentz networks. More specifically,
\ben
\item For Poincar\'e models, $R_{max}$ caps the maximum radius within the Poincar\'e disk ($R_{max}\leq  1$). If not specified, $R_{max}=1$ by default. 
\item For Lorentz models,  $L_{max}$ acts as a spatial clamp on the spatial components of the Lorentz vector (with the time component determined automatically by the Lorentz hyperboloid constraint $\langle x, x\rangle=-1$). If not specified, $L_{max} \rightarrow \infty$ by default, since the Lorentz hyperboloid has no boundary. 
\enn
Crucially, it must be noted that $R_{max}$  and $L_{max}$  do not alter the intrinsic metric or curvature of the hyperbolic space; rather, their primary function is numerical regularization. Because hyperbolic manifolds exhibit exponential volume growth, unconstrained vectors can drift toward the boundary of the Poincar\'e disk or to infinity on the unbounded Lorentz hyperboloid, leading to gradient explosion or floating point instability during VMC optimization.  Whether these bounds are set to their default unconstrained values (e.g., $R_{\text{max}} = 1.0$) or tightened to smaller spatial clamps, the underlying manifold remains strictly hyperbolic throughout optimization.
\item 
Capping $R_{\text{max}}$ or $L_{\text{max}}$ involves a practical trade-off - setting the bound too low restricts the neural network's expressivity, while setting it too high causes numerical divergence. Because each VMC optimization run for a $12 \times 12$ lattice requires up to 20 hours of compute time, conducting an exhaustive, fine-grained grid search across all combinations was computationally prohibitive given our computing resource constraints. For this proof-of-concept study, we perform focused iterative sweeps over discrete spatial bounds while holding all standard optimization hyperparameters—such as learning rates, hidden dimensions ($d_h$), sample counts, and optimizer steps—identical across both Euclidean and hyperbolic baselines. Euclidean models are evaluated over equivalent grid spaces for standard hyperparameters to ensure they are given equal tuning capacity and parameter expressivity.
\bei
\item For 1D NQS models, the best $L_{max}$ value for Lorentz RNN/GRU NQS is chosen from the list $L_{max} \in [2.0, 4.0, 6.0]$, while the best $R_{max}$ value for Poincar\'e RNN/GRU NQS is chosen from $R_{max}\in [0.65, 0.8, 1.0]$. 
\item  For 2D NQS models, we use a trial-and-error method to select the two best values of $L_{max}$ per experiment, starting with the baseline value of $L_{max}=2.0$. Depending on the magnetic field strength $B_x$, which determines the physics of the system, $L_{max}$ is either increased or decreased. For $B_x=3.0$, $L_{max}\in [2.0, 5.0]$, for $B_x=4.0$, $L_{max}\in [2.0, 1.5]$, for $B_x=2.0$, $L_{max}\in [2.0, 6.0]$. 
\eni
\eni
\item \textit{Number of NQS random seeds}: For every single NQS ansatz, three distinct NQS random seeds (111, 222, 333) corresponding to three distinct neural network random weight initializations were used to ensure variability across independent optimization runs. While evaluating a larger ensemble of seeds (e.g., 5 or larger) would be ideal, given the large number of VMC experiments and the fact that a single 2D hyperbolic NQS VMC optimization runtime on larger lattices can take up to 20 hours of execution time or more, it was computationally prohibitive (within our current computing resources) to go beyond the current number of three NQS seeds per ansatz. 

\item  \textit{Optimizer and training setup}: 
\bei
\item \texttt{Adam} optimizer is used for all Euclidean parameters (of Euclidean networks and hyperbolic networks). 
\item For Poincar\'e networks, \texttt{Riemannian SGD} optimizer \cite{ganea-1805} is used on the hyperbolic parameters.
\item For Lorentz networks, \texttt{Adam} optimizer is used on the hyperbolic parameters as well, since these are actually exponentiated Euclidean parameters (i.e. these parameters are originally Euclidean parameters that are mapped to the Lorentz hyperboloid via the exponential map Eq.(\ref{eq-l-expmap})).
\item 
The scheduled learning rate (LR) decay \texttt{ReduceLROnPlateau}\footnote{\href{https://docs.pytorch.org/docs/2.14/generated/torch.optim.lr_scheduler.ReduceLROnPlateau.html}{https://docs.pytorch.org/docs/2.14/generated/torch.optim.lr\_scheduler.ReduceLROnPlateau.html} (Last accessed October 2026.)} that automatically reduces the learning rate by half when an energy plateau is detected within 40 epochs is used in all experiments. For Euclidean parameters, the initial LR is set at $10^{-2}$ while for hyperbolic parameters, the initial LR is set at $5\times 10^{-3}$. It is a common practice, as documented in NLP works \cite{ganea-1805}), to set the hyperbolic LR smaller than the Euclidean LR. This is because the optimization landscapes of hyperbolic neural networks are much steeper and more difficult for an optimizer to navigate compared to that of a Euclidean NQS, due to the constrast between the exponential growth in volume coupled with the constant negative curvature of hyperbolic space versus the flat and zero curvature of Euclidean space.
\item During training, the saving of a model's weights is done contingent on certain strict criteria being met regarding the improvement of the network. These criteria, consisting of the lowering of the mean energy reachable by the network as well as the energy variance being under a specified tolerance threshold, are built directly in the training code implementation to automatically save the model weights as training progresses\footnote{This point is discussed in detail in \cite{hld-hypnqs-25}.}. 
\item An early stopping mechanism is also built into the training to automatically stop it if no improvement in terms of mean energy is detected within 200 epochs. As such, the maximum numbers of epochs varied for different VMC runs.
\eni
\item \textit{Further remarks}: 
\bei
\item Besides the two additional hyperparameters (spatial constraint parameter and hyperbolic learning rate) required for hyperbolic networks, we emphasize that all other training parameters are identical for Euclidean and hyperbolic networks. Additionally, it is important to note that within the same VMC experiment, the Euclidean and hyperbolic networks used always have the exact same number of trainable parameters. The additional hyperparameters of hyperbolic networks are not trainable parameters - They are fixed before training starts. 
\item Finally, we note that because of our limited computational resources, in order to run a large number of VMC experiments, the sizes of all neural networks used in this work are relatively small, and on the order of a few thousand (up to approximately $1.6\times 10^4$) parameters only. As the fundamental nature of this and our earlier works (\cite{hld-hypnqs-25}, \cite{hld-hypnqs-26}) is proof-of-concept, our focus is on show-casing the capability of our newly constructed hyperbolic NQS in a qualitative manner in comparison with their Euclidean counterparts. With more computational resources, the problem of scaling up the hyperbolic NQS introduced in our works can be readily tackled. 
\eni
\eni
The final parameter settings for the 2DTFIM experiments using 2D ansatzes (Euclidean/Lorentz 2DRNN) are listed in Table \ref{tfim2d-hyperparams} while the final parameter settings for the unrolled 2DTFIM using 1D ansatzes (Euclidean/Poincar\'e/Lorentz RNN/GRU) are listed in Table \ref{tfim2d-ur-hyperparams}. 
\begin{table}[!ht]
\centering
\begin{tabular}{cccc}
\hline\hline
$\begin{array}{c}\text{Experiment}\\ \lf[(N,N), B_x\rr] \end{array}$ 
& $\begin{array}{c}\text{2DRNN}\\\text{Ansatz}\end{array}$ 
& $\begin{array}{c} \text{Hyperparameters} \\ \text{(Parameters, initial LRs)}\end{array}$ 
& $\begin{array}{c} \text{ NQS seeds (111, 222, 333)}\\ \text{Average training per seed}\end{array}$\\
\hline\hline
$\begin{array}{l} (8,8) \\ B_x=3.0 \end{array}$ & 
$\begin{array}{l} \text{Euclidean} \\\\ \text{Lorentz} \\(L_{max}=2.0)\end{array}$ 
& $\begin{array}{l} \\\lf(5352, \lf[10^{-2}\rr]\rr) \\\\ \lf(5352, \lf[10^{-2}, 5\times 10^{-3}\rr]\rr) \\\\ \end{array}$
& $\begin{array}{c} 0.065\,\,\text{hr} \\\\ 4.3\,\,\text{hrs}\end{array}$  \\
\hline
$\begin{array}{l} (10,10) \\ B_x=3.0 \end{array}$ & 
$\begin{array}{l} \text{Euclidean}\\ \\ \text{Lorentz} \\ (L_{max}=2.0)\end{array}$ 
& $\begin{array}{l} \\\lf(7622, \lf[10^{-2}\rr]\rr) \\\\ \lf(7622, \lf[10^{-2}, 5\times 10^{-3}\rr]\rr) \\\\ \end{array}$
& $\begin{array}{c} 0.2\,\,\text{hr} \\\\ 17.2\,\,\text{hrs}\end{array}$  \\
\hline
$\begin{array}{l} (12,12) \\ B_x=2.0 \end{array}$ & 
$\begin{array}{l} \text{Euclidean} \\\\ \text{Lorentz} \\(L_{max}=2.0) \\ \\ \text{Lorentz} \\ (L_{max}=6.0)\end{array}$ 
& $\begin{array}{l} \lf(7622, \lf[10^{-2}\rr]\rr) \\\\ \lf(7622, \lf[10^{-2}, 5\times 10^{-3}\rr]\rr)\\\\\\ \lf(7622, \lf[10^{-2}, 5\times 10^{-3}\rr]\rr) \\\\ \end{array}$
& $\begin{array}{c}  0.94\,\,\text{hr}\\\\ 16.4\,\,\text{hrs}\\\\ \\21.3\,\,\text{hrs}\end{array}$  \\
\hline
$\begin{array}{l} (12,12) \\ B_x=3.0 \end{array}$ & 
$\begin{array}{l} \text{Euclidean}\\ \\ \text{Lorentz} \\ (L_{max}=2.0) \\ \\ \text{Lorentz} \\ (L_{max}=5.0)\end{array}$ 
& $\begin{array}{l} \lf(7622, \lf[10^{-2}\rr]\rr) \\\\ \lf(7622, \lf[10^{-2}, 5\times 10^{-3}\rr]\rr)\\\\\\ \lf(7622, \lf[10^{-2}, 5\times 10^{-3}\rr]\rr) \\\\ \end{array}$
& $\begin{array}{c} 0.65\,\,\text{hr}\\\\ 27.9\,\,\text{hrs}\\\\\\ 19.2\,\,\text{hrs}\end{array}$  \\
\hline
$\begin{array}{l} (12,12) \\ B_x=4.0 \end{array}$ & 
$\begin{array}{l} \text{Euclidean} \\ \\\text{Lorentz} \\ (L_{max}=2.0) \\ \\ \text{Lorentz} \\ (L_{max}=1.5)\end{array}$ 
& $\begin{array}{l} \lf(7622, \lf[10^{-2}\rr]\rr) \\\\ \lf(7622, \lf[10^{-2}, 5\times 10^{-3}\rr]\rr)\\\\\\ \lf(7622, \lf[10^{-2}, 5\times 10^{-3}\rr]\rr) \\\\ \end{array}$
& $\begin{array}{c} 0.48\,\,\text{hr}\\\\ 19.2\,\,\text{hrs}\\\\ \\16.8\,\,\text{hrs}\end{array}$  \\
\hline
\end{tabular}
\caption{Hyperparameters for the 2DTFIM VMC experiments using 2D Euclidean/Lorentz NQS ansatzes where each experiment is identified by the combination $[(N,N), B_x]$ of the square lattice size $(N,N)$ and the transverse magnetic field strength $B_x$. There are two types of ansatzes: Euclidean 2DRNN and Lorentz 2DRNN, but each Lorentz 2DRNN with a different spatial constraint hyperparameter $L_{max}$ is counted as a distinct ansatz since Lorentz 2DRNN  ansatzes with different $L_{max}$ have different levels of expressivity (as such, for those VMC experiments where $(N,N)=(12,12)$, there are 3 different ansatzes as listed in the table above). $L_{max}$ is selected by trial and error, starting with a baseline value of $L_{max}=2.0$ for all experiments. As the training time of Lorentz 2DRNN at $(N,N)=(12,12)$ is approximately 20 hours, it is not computationally possible to systematically scan for different values of $L_{max}$. For each ansatz, three different NQS seeds (111, 222, 333) are used where each NQS seed corresponds to a distinct random weight initialization parameter or a distinct VMC optimization trajectory. 
The hyperparameters listed include the number of model parameters and the two types of learning rates (LRs): Euclidean LR and hyperbolic LR. Both Euclidean and hyperbolic neural networks have Euclidean parameters that are subjected to Euclidean LR. Only hyperbolic neural networks have hyperbolic parameters that are subjected to hyperbolic LR. As such in the table above, there is only one entry for LR corresponding to Euclidean 2DRNN while there are two entries for LRs corresponding to Lorentz 2DRNN. Finally, note that the LRs listed are initial LRs that are automatically reduced by a factor of 0.5 whenever a plateau occurs in the training.  The average training time per seed, where training was done on CPU-only machines with 8-16Gb of RAM, is the time averaged over 3 separate VMC runs per ansatz.} \label{tfim2d-hyperparams}
\end{table}
\clearpage
\begin{table}[!ht]
\centering
\begin{tabular}{cccc}
\hline\hline
$\begin{array}{c}\text{Experiment}\\ \lf[(N,N)\rr] \end{array}$  
& Ansatz 
& $\begin{array}{c} \text{Hyperparameters} \\ \text{(Parameters, SC, initial LRs)}\end{array}$  
& $\begin{array}{c} \text{ NQS seeds (111, 222, 333)}\\ \text{Average training per seed}\end{array}$\\
\hline\hline
$\begin{array}{l} (8,8) \end{array}$   
& $\begin{array}{l} \text{Euclidean RNN} \\\\\\\text{Poincar\'e RNN}\\\\\\ \text{Lorentz RNN}\\\\\\
\text{Euclidean GRU} \\\\\\ \text{Poincar\'e GRU} \\\\\\\text{Lorentz GRU} \end{array}$  
& $\begin{array}{l} \lf(3902, \,\text{NA},\, 10^{-2}\rr)\\\\ 
\lf(3902, R_{max}=1.0\rr)\\ \lf[10^{-2}, 5\times 10^{-3}\rr]
\\\\\lf(3902, L_{max}=6.0\rr)\\ \lf[10^{-2}, 5\times 10^{-3}\rr]
\\\\\lf(11462, \,\text{NA}, \,10^{-2}\rr)\\\\ \lf(11462, R_{max}=1.0\rr)\\ \lf[10^{-2}, 5\times 10^{-3}\rr]
\\\\ \lf(11462, L_{max}=6.0\rr)\\ \lf[10^{-2}, 5\times 10^{-3}\rr]
\end{array}$ 
&  $\begin{array}{c} 0.06\,\,\text{hr}\\\\\\ 1.3\,\,\text{hrs}\\\\ \\2.3\,\,\text{hrs}\\\\\\
0.17\,\,\text{hr}\\\\\\  2.2\,\,\text{hrs}\\\\\\ \text{9.2\,\,\text{hrs}} \end{array}$  \\
 \hline
$\begin{array}{l} (10,10) \end{array}$   
& $\begin{array}{l} \text{Euclidean RNN} \\\\\\\text{Poincar\'e RNN}\\\\\\ \text{Lorentz RNN}\\\\\\
\text{Euclidean GRU} \\\\\\ \text{Poincar\'e GRU} \\\\\\\text{Lorentz GRU} \end{array}$  
& $\begin{array}{l} \lf(5252, \,\text{NA},\, 10^{-2}\rr)\\\\ 
\lf(5252, R_{max}=0.65\rr)\\ \lf[10^{-2}, 5\times 10^{-3}\rr]
\\\\\lf(5252, L_{max}=6.0\rr)\\ \lf[10^{-2}, 5\times 10^{-3}\rr]
\\\\\lf(15472, \,\text{NA}, \,10^{-2}\rr)\\\\ \lf(15472, R_{max}=1.0\rr)\\ \lf[10^{-2}, 5\times 10^{-3}\rr]
\\\\ \lf(15472, L_{max}=4.0\rr)\\ \lf[10^{-2}, 5\times 10^{-3}\rr]
\end{array}$ 
&  $\begin{array}{c} 0.35 \,\,\text{hr}\\\\\\ 5.2 \,\,\text{hrs}\\\\ \\12.1 \,\,\text{hrs}\\\\\\
 0.8 \,\,\text{hr}\\\\ \\ 7.0\,\,\text{hrs}\\\\ \\20.0\,\,\text{hrs} \end{array}$  \\
 \hline
$\begin{array}{l} (12,12) \end{array}$   
& $\begin{array}{l} \text{Euclidean RNN} \\\\\\\text{Poincar\'e RNN}\\\\\\ \text{Lorentz RNN}\\\\\\
\text{Euclidean GRU} \\\\\\ \text{Poincar\'e GRU} \\\\\\\text{Lorentz GRU} \end{array}$  
& $\begin{array}{l} \lf(5252, \,\text{NA},\, 10^{-2}\rr)\\\\ 
\lf(5252, R_{max}=0.8\rr)\\ \lf[10^{-2}, 5\times 10^{-3}\rr]
\\\\\lf(5252, L_{max}=4.0\rr)\\ \lf[10^{-2}, 5\times 10^{-3}\rr]
\\\\\lf(15472, \,\text{NA}, \,10^{-2}\rr)\\\\ \lf(15472, R_{max}=1.0\rr)\\ \lf[10^{-2}, 5\times 10^{-3}\rr]
\\\\ \lf(15472, L_{max}=4.0\rr)\\ \lf[10^{-2}, 5\times 10^{-3}\rr]
\end{array}$ 
&  $\begin{array}{c} 0.36\,\,\text{hr}\\\\\\ 6.3\,\,\text{hrs}\\\\\\  13.2\,\,\text{hrs}\\\\\\
0.9\,\,\text{hr}\\\\\\  22.1 \,\,\text{hrs}\\\\ \\51.8\,\,\text{hrs} \end{array}$  \\
\hline
\end{tabular}
\caption{Hyperparameters for the unrolled 2DTFIM VMC experiments with 1D NQS ansatzes where each experiment is identified by the lattice size $(N,N)$ with the magnetic field $B_x$ is fixed at $B_x=3.0$. SC stands for `spatial constraint': $R_{max}$ for Poincar\'e, $L_{max}$ for Lorentz networks, and NA (not applicable) for Euclidean models. The final $L_{max}$ value for each Lorentz NQS was chosen from [2.0, 4.0, 6.0], while the $R_{max}$ for each Poincar\'e NQS was chosen from [0.65, 0.8, 1.0]. LRs refer to the initial Euclidean and hyperbolic learning rates. Only hyperbolic networks have both types of LRs. For each ansatz, 3 distinct NQS seeds corresponding  to 3 distinct optimization trajectories were used to run the VMC experiments. The average training time per seed, where training was done on CPU-only machines with 8-16Gb of RAM, is the time averaged over 3 separate VMC runs per ansatz.} \label{tfim2d-ur-hyperparams}
\end{table}

\end{document}